\documentclass[iop, numberedappendix]{emulateapj}
\usepackage{amssymb}
\usepackage{amsmath}
\usepackage{amsfonts}
\usepackage{float}
\usepackage{relsize}
\usepackage{verbatim}
\usepackage{color}
\usepackage{comment}
\usepackage{graphicx}
\usepackage{bm}

\newcommand{\todoblank}[1]{}

\newcommand{\added}[1]{{#1}}
\newcommand{\addedA}[1]{{#1}}
\newcommand{\addedB}[1]{{#1}}
\newcommand{\addedC}[1]{{#1}}

\graphicspath{{figures/}}

\def\be{\begin{equation}}
\def\ee{\end{equation}}
\def\ba{\begin{eqnarray}}
\def\ea{\end{eqnarray}}

\newcommand{\DMhat}{\widehat{\mathrm{DM}}}

\newcommand{\DM}{\ensuremath{\mathrm{DM}}} 
\newcommand{\SN}{\ensuremath{\mathrm{S/N}}}

\newcommand{\J}{\ensuremath{\mathrm{J}}}
\newcommand{\W}{\ensuremath{\mathrm{W}}}

\newcommand{\DISS}{\ensuremath{\mathrm{DISS}}}

\newcommand{\PBF}{\ensuremath{\mathrm{PBF}}}

\newcommand{\Nphi}{\ensuremath{N_\phi}}

\newcommand{\Weff}{\ensuremath{W_\mathrm{eff}}}

\newcommand{\tel}{\ensuremath{\mathrm{tel}}}
\newcommand{\Ae}{\ensuremath{A_\mathrm{e}}}

\newcommand{\Tsys}{\ensuremath{T_{\mathrm{sys}}}}

\newcommand{\niss}{n_{\rm ISS}}

\newcommand{\Dtd}{\Delta t_{\mathrm{d}}}
\newcommand{\Dtdo}{\Delta t_{\mathrm{d,0}}}
\newcommand{\Dnud}{\Delta \nu_{\mathrm{d}}}
\newcommand{\Dnudo}{\Delta \nu_{\mathrm{d,0}}}

\newcommand{\taud}{\tau_{\mathrm{d}}}
\newcommand{\taudo}{\tau_{\mathrm{d,0}}}
\newcommand{\tablespacer}{\ensuremath{\vphantom{\displaystyle{\left(\frac{\nu}{\nu_0}\right)^-22/5}}}}

\newcommand{\Tmat}{\mathbf{T}}
\newcommand{\Dmat}{\mathbf{D}}
\newcommand{\Xmat}{\mathbf{X}}
\newcommand{\XmatT}{\mathbf{X^T}}
\newcommand{\Umat}{\mathbf{U}}
\newcommand{\UmatT}{\mathbf{U^T}}
\newcommand{\Pmat}{\mathbf{P}}
\newcommand{\Cmat}{\mathbf{C}}
\newcommand{\CEmat}{\mathbf{C_E}}
\newcommand{\Jmat}{\mathbf{J}}
\newcommand{\Smat}{\mathbf{S}}
\newcommand{\Cmatinv}{\mathbf{C^{-1}}}
\newcommand{\thetamat}{\bm{\theta}}
\newcommand{\epsilonmat}{\bm{\epsilon}}

\newcommand{\approxprop}{\mathrel{\vcenter{
  \offinterlineskip\halign{\hfil$##$\cr
    \propto\cr\noalign{\kern2pt}\sim\cr\noalign{\kern-2pt}}}}}

\shorttitle{Optimal Frequencies for Radio Pulsars}
\shortauthors{Lam et al.}

\begin{document}

\title{
Optimal Frequency Ranges for Sub-Microsecond Precision Pulsar Timing
}
\author{ 
M.\,T.\,Lam\altaffilmark{1,2},
M.\,A.\,McLaughlin\altaffilmark{1,2},
J.\,M.\,Cordes\altaffilmark{3},
S.\,Chatterjee\altaffilmark{3},
T.\,J.\,W.\,Lazio\altaffilmark{4}
}
\altaffiltext{1}{Department of Physics and Astronomy, West Virginia University, White Hall, Morgantown, WV 26506, USA; michael.lam@mail.wvu.edu}
\altaffiltext{2}{Center for Gravitational Waves and Cosmology, West Virginia University, Chestnut Ridge Research Building, Morgantown, WV 26505}
\altaffiltext{3}{Department of Astronomy and Cornell Center for Astrophysics and Planetary Science, Cornell University, Ithaca, NY 14853, USA}
\altaffiltext{4}{Jet Propulsion Laboratory, California Institute of Technology, 4800 Oak Grove Dr. Pasadena CA 91109, USA}

\begin{abstract}

Precision pulsar timing requires optimization against measurement errors and  astrophysical variance from the neutron stars themselves and the interstellar medium. We investigate optimization of arrival time precision as a function of radio frequency and bandwidth. We find that increases in bandwidth that reduce the contribution from receiver noise are countered by the strong chromatic dependence of interstellar effects and intrinsic pulse-profile evolution. The resulting optimal frequency range is therefore telescope and pulsar dependent. We demonstrate the results for five pulsars included in current pulsar timing arrays and determine that they are not optimally observed at current center frequencies. \addedB{For those objects, we find that better choices of total bandwidth as well as center frequency can improve the arrival-time precision.} Wideband receivers centered at \addedC{somewhat higher} frequencies \addedC{with respect to the currently adopted receivers} can reduce required overall integration times and provide significant improvements in arrival time uncertainty by a factor of $\sim\!\!\sqrt{2}$ in most cases, assuming a fixed integration time. We also discuss how timing programs can be extended to pulsars with larger dispersion measures through the use of higher-frequency observations.



\end{abstract}


\section{Introduction}

Pulsars have been used in some of the most constraining tests of gravity and general relativity \citep[see][for an overview]{Will2014}, as laboratories for super-dense nuclear equation-of-state experiments \added{\citep{Demorest+2010,lp2016}}, and as detectors of low-frequency (nanohertz to microhertz) gravitational waves \citep{PPTACW,PPTABWM,EPTAGWB,NG5BWM,Shannon+2015,Lasky+2016,NG9GWB,EPTACW}. Observations of pulse times of arrival (TOAs) allow us to develop a timing model that tracks every rotational phase of the pulsar \citep{Verbiest+2009,cs2010}. Advancing pulsar timing tests will require the highest arrival-time precision possible \citep{IPTADR1noise,NG9EN}.

Many frequency-dependent effects distort pulsar signals along the line of sight \citep{Cordes2013,Stinebring2013}. Intrinsic pulse emission varies as a function of frequency \citep[e.g.,][]{pdr2014,Liu+2014}. The interstellar medium (ISM) causes both time and frequency-dependent pulse shape and flux changes as well as arrival time delays from numerous optical effects \citep{Palliyaguru+2015,Levin+2016,NG9WN}. Measurement at the Earth is limited to specific receiver instrumentation and frequency ranges, as well as from observation through specific radio-frequency-interference (RFI) environments \citep[e.g.,][]{IPTADR1}. The TOAs and their uncertainties, and therefore the ultimate achievable timing precision, will be affected by the choice of radio frequencies with which we observe.

\added{This paper considers optimization of timing precision as a function of observing frequency and bandwidth. Our results are generally applicable to any high-timing-precision experiment with pulsars. We apply our results to several pulsars observed by the North American Nanohertz Observatory for Gravitational Waves (NANOGrav; \citealt{McLaughlin2013}) collaboration. \addedA{NANOGrav observes many pulsars with frequencies ranging from $\sim$0.3$-$2.5~GHz but our analysis is focused on a select few.}}

In~\S\ref{sec:FD_effects}, we describe the various frequency-dependent effects in pulsar timing and develop the methodology to compute a single TOA uncertainty. Table~\ref{table:effects} provides a summary of the different effects we discuss and the general forms of the equations. We discuss implications for specific high-precision millisecond pulsars (MSPs) observed by NANOGrav \addedB{(hereafter ``NANOGrav pulsars'' for short)} in \S\ref{sec:NANOGravMSPs} and then expand our analysis to other MSPs broadly in \S\ref{sec:broadimplications}. We conclude in \S\ref{sec:discussion}.



\begin{center}
\begin{deluxetable*}{lc|cc}
\tablecolumns{4}
\tablecaption{\addedB{Selected} Timing Effects}
\tablehead{
\colhead{Term} & \colhead{Symbol} & \colhead{Dependence$^{\rm a}$} & \colhead{Section Discussed}\\
}
\startdata
Template-Fitting & $\sigma_{\SN}$ & $\displaystyle{\frac{\Weff(\nu,\taud)}{S(\nu,\taud)\sqrt{\Nphi}} \approxprop \sqrt{BT}}$ & \S\ref{sec:template-fitting} \\
\hspace{3ex} Flux-Density Spectrum & $I$ & $\displaystyle{I_0\left(\frac{\nu}{\nu_0}\right)^{\alpha}}$ & \S\ref{sec:template-fitting}, Eq.~\ref{eq:Inu}\\
\hspace{3ex} \addedB{Intrinsic} Profile Evolution & $U_{\rm int}$ & Varies\tablespacer & \S\ref{sec:template-fitting}, Eq.~\ref{eq:Uobs}\\
\hspace{3ex} Pulse Broadening & $h_{\PBF}$ & $\displaystyle{\taudo\left(\frac{\nu}{\nu_0}\right)^{-22/5}}$ & \S\ref{sec:template-fitting}, Eq.~\ref{eq:Uobs}\\
\hspace{3ex} System Temperature & $T_{\rm sys}$ & $\cdots$\tablespacer & \S\ref{sec:template-fitting}, Eq.~\ref{eq:Tsys} \\
\hspace{6ex} Cosmic Microwave Background & $T_{\rm CMB}$ & Constant\tablespacer & \S\ref{sec:template-fitting}, Eq.~\ref{eq:Tsys} \\
\hspace{6ex} Receiver Bandpass & $T_{\rm rcvr}$ & $\sim$Constant\tablespacer & \S\ref{sec:template-fitting}, Eq.~\ref{eq:Tsys} \\
\hspace{6ex} Galactic Background & $T_{\rm Gal}$ & $\displaystyle{T_{\rm Gal,0}\left(\frac{\nu}{\nu_0}\right)^{-\beta}}$ & \S\ref{sec:template-fitting}, Eq.~\ref{eq:Tsys},\ref{eq:Tgal} \\
Pulse Phase Jitter & $\sigma_\J$ & $\sim$Constant\tablespacer & \S\ref{sec:jitter} \\
Diffractive Interstellar Scintillation$^{\rm b}$ & $\sigma_{\DISS}$ & $\displaystyle{\approx \taudo\left(\frac{\nu}{\nu_0}\right)^{-8/5} \left(\frac{\Dtdo \Dnudo}{\eta_t \eta_\nu T B}\right)^{1/2}}$ & \S\ref{sec:DISS} \\ 
DM Estimation & $\sigma_{\delta\DM}$ & $\cdots$\tablespacer & \S\ref{sec:DMestimation}, Eq.~\ref{eq:sigma_deltaDM}\\
\hspace{3ex} from white-noise$^{\rm c}$ & $\sigma_{\DMhat}$ & $\displaystyle{\simeq \frac{\epsilon_{\nu_1} - r^2 \epsilon_{\nu_2}}{r^2 - 1}}$\tablespacer & \S\ref{sec:DMestimation}\\
\hspace{3ex} from Systematic Chromatic Delays$^{\rm c}$ & $\sigma_{\delta t_C}$ & $\displaystyle{\simeq \frac{t_{C,\nu_1} - r^2 t_{C,\nu_2}}{r^2 - 1}}$\tablespacer & \S\ref{sec:DMestimation}\\
\hspace{3ex} from Frequency-Dependent DM & $\sigma_{\DM(\nu)}$ & $\displaystyle{\approx 9~\mathrm{ns}~E_{11/3}(r) \left(\frac{\nu}{\mathrm{GHz}}\right)^{-1} \left(\frac{\nu/\Dnud(\nu)}{100}\right)^{5/6}}$ & \S\ref{sec:DMestimation}, Eq.~\ref{eq:sigma_DMnu}\\
Telescope & $\sigma_\tel$ & $\cdots$\tablespacer & \S\ref{sec:telescope} \\
\hspace{3ex} Polarization Calibration & $\sigma_{\rm pol}$ & $\displaystyle{\varepsilon \pi_{\rm V} W \sim \varepsilon \eta^{1/2} \pi_{\rm L} W}$ & \S\ref{sec:telescope}, Eq.~\ref{eq:sigma_pol}
\enddata
\footnotetext{All variables are discussed in text.}
\footnotetext{When the number of scintles $\niss$ is large in both time and frequency.}
\footnotetext{The form here shows the scaling for two individual narrowband frequencies.}
\label{table:effects}
\end{deluxetable*}
\end{center}

\section{Frequency-Dependent TOA Uncertainty}
\label{sec:FD_effects}

In this section, we discuss the various components contributing to pulse TOA uncertainty as a function of frequency. Our goal is to compute the TOAs and uncertainties referenced to the infinite-frequency arrival time, $t_\infty$. For frequencies $\nu_1$ and $\nu_2$ with $\nu_1 < \nu_2$, we define the bandwidth $B = \nu_2 - \nu_1$ and where convenient the frequency ratio $r = \nu_2/\nu_1$. The observing timespan will be given by $T$. Throughout the paper, we use the subscript `0' to denote a parameter referenced to a center frequency of $\nu_0$ unless otherwise specified.

\subsection{Timing and Noise Model}

Our approach in this paper is to examine the frequency-dependent timing and noise model we use to describe our TOAs and then formulate an uncertainty on $t_\infty$. Similar to \citet{cs2010} and \citet{css2016}, we consider our timing model for frequency-dependent TOAs $t_\nu$ to be
\be
t_\nu = t_\infty + \frac{K \DM(\nu)}{\nu^2} + t_{C,\nu} + \epsilon_\nu,
\label{eq:timing_model}
\ee
where $K\DM(\nu)/\nu^2$ is the dispersive delay given by a frequency-dependent dispersion measure (DM) with dispersion constant $K \approx 4.149~\mathrm{ms~GHz^2~pc^{-1}~cm^3}$ \citep{handbook,css2016}, $t_{C,\nu}$ represents an additional systematic chromatic delay term, and $\epsilon_\nu$ is the TOA uncertainty measured at specific frequency $\nu$. In general, $\epsilon_\nu$ will contain components from multiple white-noise sources described by a covariance matrix $\Cmat_{\nu\nu'}$\added{, where $\nu$ and $\nu'$ denote two separate observing sub-bands}. The three white-noise (i.e., uncorrelated, for this case considered in time) components of our covariance matrix are additive, i.e.,
\be
\Cmat_{\nu\nu'} = \Smat_{\nu\nu'} + \Jmat_{\nu\nu'} + \Dmat_{\nu\nu'},
\ee
where $\Smat_{\nu\nu'}$ is from template fitting, $\Jmat_{\nu\nu'}$ is from pulse jitter, and $\Dmat_{\nu\nu'}$ is from diffractive interstellar scintillations (DISS). Template-fitting errors are calculated from a matched-filtering procedure using a template shape compared with a data profile containing additive noise \added{(such as from radiometer noise)} that varies in both time and frequency \added{\citep[see][for more details]{NG9WN}}. Pulse jitter results from individual pulses varying stochastically, breaking the matched-filtering assumption of the averaged pulse profile being an exact copy of the template plus additive noise, and thus is uncorrelated between pulses in time though does have some correlation in frequency. For DISS, the stochastically varying ISM along the line of sight similarly results in observed pulses deviating from the template shape which are uncorrelated in time and frequency beyond the characteristic scintillation time and frequency scales \added{\citep{Levin+2016}}. All three effects will be discussed in more detail in later subsections \added{(\S\ref{sec:template-fitting}, \S\ref{sec:jitter}, \S\ref{sec:DISS}, respectively)}. Multiple effects can cause TOA uncertainties that are correlated over longer timescales than a single epoch though we do not discuss them here in this paper; \added{errors on any timing parameters will typically be small due to the long-term timing fit.}

We will use the combined covariance matrix to determine the white-noise uncertainty averaged over the single observation. Additional error terms will be added in quadrature to this single white-noise uncertainty. We will describe each of the noise components in detail before discussing the DM fitting procedure. For brevity, we will drop the subscripts on all matrices. 


Rather than deal with $N$ number of independently-measured pulses across frequency channels over our bandpass and therefore $N$ template-fitting uncertainties, we choose to combine the errors into a single number, thus providing a metric to compare TOA uncertainties as a function of center frequency and bandwidth. We compute the single white-noise uncertainty component on $t_\infty$ following the procedure in \citet{NG5BWM},
\be
\CEmat = \mathbf{\left(\UmatT\Cmatinv\Umat\right)^{-1}},
\label{eq:epoch_average}
\ee
where the matrix $\Umat$ is the ``exploder matrix'' defined in \citet{NG5CW} that in general groups the multifrequency arrival times taken over many observations into specific epochs and for this work specifically groups the frequency-dependent arrival times into a single epoch by setting $\Umat = \mathrm{col}(1)$, a column matrix where each element is 1. Again, the covariance matrix $\Cmat$ is the sum of the three separate covariance matrices $\Smat$, $\Jmat$, and $\Dmat$. The white-noise uncertainty $\sigma_\W$ is then the square root of the single element of $\CEmat$.

We define the total TOA uncertainty (squared) on $t_\infty$ as
\be
\sigma_{\rm TOA}^2 = \sigma_\W^2 + \sigma_{\delta \DM}^2 + \sigma_{\rm tel}^2
\label{eq:sigma_TOA}
\ee
where the variances in order are the squares of: the white-noise errors, the DM estimation error, and the telescope error. We wish to reinforce that $\sigma_{\rm TOA}$ is separate from $\epsilon_\nu$, which is the per-frequency uncertainty on $t_\nu$. We will describe the three components in depth in the following subsections.

\subsection{White-Noise Variations}

The three components of the white-noise TOA uncertainties are given by
\be
\sigma_\W^2 = \sigma_\SN^2 + \sigma_\J^2 + \sigma_\DISS^2,
\label{eq:sigma_W}
\ee
where $\sigma_\SN$ is the template-fitting error, $\sigma_\J$ is the jitter error, and $\sigma_\DISS$ is the scintillation error. All three are discussed in detail in \citet{NG9WN}. While the three terms have been shown to be uncorrelated between epochs (and thus are white noise), $\sigma_\J$ is correlated in frequency and $\sigma_\DISS$ has a characteristic time and frequency correlation structure, discussed shortly.

\subsubsection{Template-Fitting Error}
\label{sec:template-fitting}

We estimate TOAs calculated with a template-fitting approach \citep{Taylor1992}. In order to calculate the template-fitting error, we assume that the pulse profile is an exact copy of a template shape with additive noise. The minimum TOA error can be calculated from the effective width of the pulse $\Weff$ and the $\SN$ (peak to off-pulse rms, written as $S$ for clarity in equations),
\be
\sigma_{\SN}(S) = \frac{\Weff}{S\sqrt{\Nphi}},
\label{eq:tf_error}
\ee
where $\Nphi$ is the number of samples (bins) across pulse phase $\phi$ from 0 to 1. Changes in $\Nphi$ will change the rms noise and thus $S$; however the product $S\Nphi^{1/2}$ is invariant \citep{NG9WN}. Use of the equation assumes that $\Weff$ is constant over the band. We note that in the low-S/N regime, the true template-fitting error is larger than given by Eq.~\ref{eq:tf_error} because the template-matching procedure begins to fail and the error becomes significantly non-Gaussian \citep[][Appendix~B]{NG9yr}. 

Since the pulse profile changes over the frequency range, the template-matching assumption fails if we are to average the pulse over the band. Pulse profiles tend to broaden at low frequencies intrinsically (though not necessarily), pulse profile components can overlap in pulse phase or appear/disappear altogether, and multipath scattering broadens pulses. Therefore, to compute the template-matching error, we assume that the pulse shape in a sufficiently narrow frequency channel is constant. The assumption has the added benefit of allowing for a single $S$ for a narrow frequency channel.

The effective width of the pulse at a given frequency can be written as 
\ba
& & \hspace{-3ex} \Weff(\nu|\taudo) = \nonumber \\
& & \hspace{-3ex} \frac{P}{\Nphi^{1/2} \!\left[\displaystyle \sum_{i=1}^{\Nphi-1}\!\!\left[U_{\rm obs}(\phi_i,\nu|\taudo)\!-\!U_{\rm obs}(\phi_{i-1},\nu|\taudo)\right]^2\!\right]^{1/2}},
\label{eq:Weff}
\ea
where $P$ is the pulsar spin period, $\taudo$ is the scattering timescale at reference frequency $\nu_0$, and the observed pulse template \addedC{$U_{\rm obs}$ scaled to a peak of unity} is equal to the intrinsic pulse shape convolved with the pulse broadening function (PBF),
\be
U_{\rm obs}(\phi|\nu,\taudo) = U_{\rm int}(\phi|\nu) \ast h_{\rm PBF}(\nu,\taudo).
\label{eq:Uobs}
\ee
\added{The differences in the \addedB{denominator} in Eq.~\ref{eq:Weff} are related to the derivative of the template and so pulses with sharp features will have a smaller $\Weff$; the effective width is not equivalent to the width of a pulse component.} Since $\taud$ will change as a function of frequency proportional to $\nu^{-22/5}$ \addedB{(see \S\ref{sec:DISS})}, the width of the PBF will also change as a function of frequency. The PBF \addedB{that conserves pulse fluence} is approximately a negative exponential which causes the pulse not only to broaden but the amplitude to decrease \added{\citep{Bhat+2003,Geyer+2017}}. Therefore, the effective width increases and the pulse S/N decreases, causing drastic changes to $\sigma_{\SN}$ at lower frequencies.

Now we consider the pulsar signal, the numerator of the signal-to-noise ratio. Above frequencies of 100~MHz, the period-averaged flux densities can be described by a simple power law,
\be
I(\nu) = I_0\left(\frac{\nu}{\nu_0}\right)^{\alpha}, 
\label{eq:Inu}
\ee
where $I_0$ is the intensity constant and $\alpha$ is the spectral index \citep{handbook}. The mean value of $\alpha = -1.60 \pm 0.03$ \added{\citep{Jankowski+2017}}.

For a pulsar signal with a power-law flux density in frequency, the \addedC{period-averaged} signal-to-noise ratio is 
\be
\bar{S}(\nu) = \frac{I_0 \left(\frac{\nu}{\nu_0}\right)^{-\alpha} }{\Tsys(\nu) / \left(\Ae/2k\right) } \sqrt{N_{\rm pol} B T/\Nphi}, 
\label{eq:SN}
\ee
where the numerator represents the pulsar signal while the denominator and factor in the square root represents the rms off-pulse noise from the radiometer equation \addedC{(the factor of $N_{\rm pol} = 2$ represents the number of polarizations)}. \addedC{In order to obtain the peak-to-off-pulse signal-to-noise ratio $S$, we must multiply Eq.~\ref{eq:SN} by a factor of} 
\ba
S(\nu) & \equiv & U_{\rm scale}(\nu) \bar{S}(\nu) \nonumber\\
& = & \frac{1.0}{\bar{U}_{\rm obs}(\nu)} \bar{S}(\nu)
\label{eq:Uscale}
\ea
\addedC{where $\bar{U}_{\rm obs}(\nu)$ is simply the mean of $U_{\rm obs}(\nu)$, calculated as $\Nphi^{-1}\sum_{i=0}^{\Nphi-1} U_{\rm obs}(\phi_i,\nu)$. The factor of 1.0 in the numerator comes from the template being scaled to peak unity as previously discussed.} \added{We write the ratio as $S$ following the convention in \citet{NG9WN} for brevity in equations}. Again, $B$ is the observing bandwidth and $T$ is the observation duration. The factor of $\Ae/2k = (\Ae / 2760~\mathrm{m}^2)~\mathrm{K/Jy}$ in the denominator is the gain that converts the system temperature ($\Tsys$) equivalent into the total measured noise power in flux density units. 

In addition to altering the effective width $\Weff$ of the pulse profiles, multipath scattering will cause pulse profiles to have lower S/N since the total flux will be broadened across pulse phase. Since the scattering timescale $\taud$ varies as a function of frequency (proportional to $\nu^{-22/5}$ for a Kolmogorov medium), the pulsar signal term in the numerator will therefore take a more complicated, frequency-dependent form, $I(\nu|\taudo)$. Free-free absorption has a negligible effect for pulsar-frequency combinations used for sub-microsecond timing precision.

The system temperature $\Tsys$ is a combination of all temperature contributions measured by the antenna, such as from the Cosmic Microwave Background (CMB; $T_{\rm CMB}$), the receiver bandpass ($T_{\rm rcvr}$), the Galactic background ($T_{\rm Gal}$), etc., and can be written as the sum of those contributions:
\ba
\Tsys(\nu) & = & T_{\rm CMB} + T_{\rm rcvr}(\nu) + T_{\rm Gal}(\nu) + \cdots \nonumber \\
& \approx & T_{\rm const} + T_{\rm Gal,0}\left(\frac{\nu}{\nu_0}\right)^{-\beta}.
\label{eq:Tsys}
\ea
The response of the receiver is also not constant over the band and will therefore also affect $S(\nu)$, however we will assume \addedC{here} that it is constant and group it together with the CMB temperature to form $T_{\rm const}$. For the frequency ranges we consider, we use the form in \citet{SKAmemo95} for the Galactic background,
\be
T_{\rm Gal}(\nu) = 20~\mathrm{K}\left(\frac{\nu}{0.408~\mathrm{GHz}}\right)^{-2.75}.
\label{eq:Tgal}
\ee
At frequencies above about 10~GHz, a thermal Galactic component can become dominant over this nonthermal component. However, for the range of frequencies that we consider here, this thermal component will always be sub-dominant for the vast majority of lines of sight and we do not consider it here. 

\added{The Galactic component will also have some Galactic latitude and longitude dependence that changes both the coefficient and the spectral index of $T_{\rm Gal}$ \citep{SKAmemo95,Kogut+2011}. In addition, the Sun will also increase $\Tsys$ as a function of time; both the intrinsic emission and the angular size on the sky will have frequency dependence \citep{Oberoi+2017}. These effects primarily dominate at the lowest frequencies we will consider and cause the template-fitting uncertainty to increase. We will show later that the optimal frequency ranges tend to be at much higher frequencies and these spatial- and time-dependent contributions will be negligible there. Therefore, we will ignore them in our analysis for simplicity though they can be considered for future analyses depending on the applications.}



Combining Eqs.~\ref{eq:tf_error}, \ref{eq:Weff} and \ref{eq:SN}, the template-fitting error can be written as
\be
\sigma_{\SN}(\nu|\taudo) = \frac{\Weff(\nu|\taudo)}{S(\nu|\taudo)\sqrt{\Nphi}},
\ee
where we have explicitly included a scattering timescale that modifies both the pulse shape and intensity. The template-fitting covariance matrix is then diagonal with components
\be
\Smat = \mathrm{diag}\left[\sigma_{\SN}^2(\nu|\taudo)\right].
\ee

\subsubsection{Pulse Phase Jitter}
\label{sec:jitter}

While average pulse profiles at a specific frequency are stable over time, the shapes of individual pulses stochastically vary \citep{cd1985}. Known as pulse jitter, a finite sum of pulses will deviate from the template slightly, breaking the template-fitting assumption of the profile being the sum of the template plus additive noise. Therefore an additional TOA uncertainty must be taken into account. Since the average pulse shapes change as a function of frequency, we expect that pulse jitter will change as a function of frequency such that when summed they yield the observed frequency-dependent pulse profile evolution; comparisons between the two regimes, however, have not been studied previously. \citet{sod+2014} found that the jitter noise decorrelates over a bandwidth of approximately 2~GHz for observations of time-averaged pulses covering $\sim$0.7$-$3.6~GHz for PSR~J0437$-$4715. \citet{NG9WN} found statistically different values of jitter noise for many MSPs in the NANOGrav pulsar timing array (PTA) as a function of receiver band observed. The values are expected to be correlated between frequency bands though an analysis has not yet been undertaken.


For narrow bands over which the rms jitter $\sigma_\J$ is constant, the covariance matrix $\Jmat$ has all components set to $\sigma_\J^2$. In general however the jitter covariance matrix is given by
\be
\Jmat = \rho_\J(\nu,\nu')\sigma_\J(\nu) \sigma_J(\nu')
\ee
where $\rho_\J$ describes the correlation of the variance between the two frequency bands and here we set $\rho = 1$ for simplicity, roughly consistent with \citet{sod+2014}.

\subsubsection{Scintillation Noise}
\label{sec:DISS}

DISS modulates the observed pulse S/N by a gain $g$ (different from the telescope gain) with a probability density function (PDF) given by
\be
f_S(S \vert \niss) = \frac{(S\niss/S_0)^{\niss}}{S\Gamma(\niss)} e^{-S\niss/S_0}\Theta(S)
\label{eq:sn_pdf}
\ee
where $S_0$ is the mean S/N, $\niss$ is the number of scintles, $\Gamma$ is the gamma function, and $\Theta$ is the Heaviside step function \citep{NG9WN}. When the number of scintles $\niss$ is large, the PDF tends towards a constant (the mean) S/N, $f_S(S\vert \niss) \rightarrow \delta(S-S_0)$. The number of scintles $\niss$ scales as 
\be
\niss \approx \left(1+\eta_t \frac{T}{\Dtd}\right)\left(1+\eta_\nu \frac{B}{\Dnud}\right),
\label{eq:niss}
\ee
where $\Dtd$ and $\Dnud$ are the scintillation bandwidth and timescale respectively, $B$ is again the total observing bandwidth, and $T$ is the observing duration. The filling factors $\eta_t, \eta_\nu$ are typically in the range 0.1 to 0.3 depending on the scattering geometry along the line of sight \citep{cs2010,Levin+2016}. For a Kolmogorov medium, the scintillation parameters scale with frequency \citep{cr98,NE2001} as
\ba
\Dtd(\nu) &=& \Dtdo \left(\frac{\nu}{\nu_0}\right)^{6/5}~\mathrm{and}\\
\Dnud(\nu) &=& \Dnudo \left(\frac{\nu}{\nu_0}\right)^{22/5}.
\ea


In addition to modulating the pulse gain, DISS causes the measured PBF to vary stochastically. Like pulse jitter, the \added{epoch-to-epoch} deviation from the template shape introduces an additional error that must be accounted for in the TOA uncertainty \added{(the finite scintle effect, see \citealt{cwd+1990} for a more thorough discussion)}. When added in quadrature to the TOA uncertainty, the scintillation noise component takes the form
\be
\sigma_{\DISS}(\nu) \approx \frac{\taud(\nu)}{\sqrt{\niss(\nu)}}
\label{eq:DISS_error}
\ee
when $\niss$ is large and $\lesssim\!\!\taud$ when there is only one scintle or a partial scintle across the band \citep{cwd+1990}. The pulse-broadening timescale $\taud$ is related to the scintillation bandwidth by $\taud = C_1 / (2\pi\Dnud)$, where $C_1$ is a coefficient of order unity. \addedB{We assumed in our analysis that the ISM is a fully-uniform Kolmogorov medium and adopted} $C_1 = 1.16$ \addedC{and $\eta_t = \eta_\nu = 0.2$} \citep{cr98,cs2010}.

The DISS covariance matrix is
\be
\Jmat = \rho_{\DISS}(\nu,\nu')\sigma_{\DISS}(\nu) \sigma_{\DISS}(\nu'),
\ee
where like $\rho_\J$, $\rho_{\DISS}$ describes the correlation of the variance between the two frequency bands though it depends on the scintillation time and frequency correlation scales at both frequencies. In the application of our analysis, we will make a simplifying assumption that $\rho_{\DISS}$ is split into two regimes, where there are multiple scintles $\niss$ in the band and therefore uncorrelated and where $\niss \approx 1$ and so $\rho_{\DISS} \rightarrow 1$. Our assumption is discussed further in \S\ref{sec:NANOGravMSPs} when we apply our analysis to specific MSPs.

\subsection{Systematic Delay Errors}

\subsubsection{\DM~Fluctuations}
\label{sec:DMestimation}

One of the largest timing errors from ISM effects comes from DM misestimation. Typically, pulsars are observed at multiple frequencies covering a wide band in order to gain infinite-frequency timing precision. We include three contributions to the DM estimation error: from additive white-noise uncertainties ($\sigma_{\DMhat}$), from frequency-dependent DM ($\sigma_{\DM(\nu)}$), and from systematic changes in the PBF ($\sigma_{\delta t_C}$). Stochastic changes from the finite-scintle effect are included in the covariance matrix whereas we let the chromatic delay term $t_{C,\nu}$ represent unknown changes in the scattering delay timescale over many epochs.

In this work, we consider only a fit for the dispersive delay, which follows most PTA procedures up to now. We follow the formalism in Appendix D of \citet{cs2010} to calculate the DM estimation error over many frequency channels, taking into account that the covariance matrix we consider is not diagonal. The timing model can be written in matrix form as $\Tmat = \Xmat \thetamat + \epsilonmat$, where $\Tmat = \mathrm{col}(t_k)$ is the column vector of arrival times for the $k$-th frequency, $\Xmat$ is the design matrix given by $\mathrm{matrix}(1~\nu_k^2)$ from a fit for only $t_\infty$ and dispersive delays, $\thetamat$ is the parameter vector $\mathrm{col}(t_\infty, K\DM)$, and $\epsilonmat$ are the errors described by the covariance matrix $\Cmat$. In the case of the design matrix we use here, we are assuming a timing model as in Eq.~\ref{eq:timing_model} but with a fit only for a constant DM over the observation, i.e., we do not fit for the additional chromatic delay since again we are only considering the model of current PTA procedures and we do not yet consider the frequency-dependence of DM \citep{css2016}. The standard error on $t_\infty$ from a fit for DM, which we denote $\sigma_{\DMhat}$, is given by the first element of the covariance matrix for the parameter errors $(\XmatT \Cmatinv \Xmat)^{\mathbf{-1}}$. In Table~\ref{table:effects}, we provide the simplified form in the case where only two individual frequencies are used. 

We model systematic variations in the PBF as a power-law scaling for the chromatic delay, $t_{C,\nu} = a_C \nu^{-X}$. \citet{fc90} assume $X = 4$ because they assume that the geometric increase in the propagation path is the greatest contribution after the dispersive delay (and the barycentric angle-of-arrival correction, see the discussion in \citealt{LamDMt}). A scaling $X = 4.4$ occurs in the moderate scattering regime for a Kolmogorov medium \citep{cs2010}, though there are known sources with indices that deviate from such a scaling (e.g., \citealt{Bhat+2004,Lohmer+2004}). Here we take $X = 4.4$. The parameter estimate for $t_\infty$ is given by the first element of $\thetamat = (\XmatT \Cmatinv \Xmat)^{\mathbf{-1}} \XmatT \Cmatinv \Tmat$. If we let $\Tmat = \Tmat_0 + \Tmat_C$, where $\Tmat_0$ describes the components of the TOAs modeled by the traditional DM fit discussed previously and $\Tmat_C$ describes the systematic $t_{C,\nu}$, then the systematic error is simply the difference between the fit when only DM is accounted for and when the chromatic term is accounted for, i.e.,
\be
\bm{\delta}\thetamat = (\XmatT \Cmatinv \Xmat)^{\mathbf{-1}} \XmatT \Cmatinv \Tmat_C.
\ee
We add this systematic error $\sigma_{\delta t_C}$ in quadrature to the standard error $\sigma_{\DMhat}$. Again, we automatically account for stochastic changes in the PBF from DISS since scintillation noise is included in $\Cmat$. Estimation and uncertainties on the shape of the profiles as a function of frequency will also enter into $\sigma_{\delta t_C}$ but we have assumed perfect knowledge of the shape changes and do not include the effect here. As with $\sigma_{\DMhat}$, we provide the simplified form of $\sigma_{\delta t_C}$ in Table~\ref{table:effects} when only two frequency channels are used.

The frequency dependence of DM is described in \citet{css2016} as a result of ray paths at different frequencies covering different volumes through the ISM. They primarily consider observations at two spot frequencies \addedC{$\nu_1$ and $\nu_2$}; observations covering greater frequency ratios $r = \nu_2/\nu_1$ will have greater error in DM estimation since the difference in DM between the two frequencies will statistically grow larger. We combine Eqs.~15 and 25 of their paper to obtain a scaling for the rms timing error \addedC{between the two frequencies} for a uniform Kolmogorov medium,
\ba
& & \hspace{-3ex}\sigma_{\DM(\nu);\mathrm{pair}}(\nu_1,\nu_2) \nonumber \\
& & \approx 9~\mathrm{ns}~ E_{11/3}(r) \left(\frac{\nu_2}{\mathrm{GHz}}\right)^{-1} \left(\frac{\nu_2/\Dnud(\nu_2)}{100}\right)^{5/6},
\label{eq:sigma_DMnu}
\ea
where $E_{11/3}(r)$ is a dimensionless quantity that contains all of the relative frequency dependence with respect to the type of medium. \addedC{The covariance matrix for the total frequency-dependent DM error $\sigma_{\DM(\nu)}$ can be constructed simply with components $\sigma_{\DM(\nu);\mathrm{pair}}(\nu,\nu^\prime)$.}



Again, we add this error in quadrature to the previously discussed DM estimation errors, and thus the total DM estimation error (squared) is
\be
\sigma_{\delta\DM}^2 = \sigma_{\DMhat}^2 + \sigma_{\delta t_C}^2 + \sigma_{\DM(\nu)}^2.
\label{eq:sigma_deltaDM}
\ee
All terms depend on the center frequency $\nu_0$ and bandwidth $B$ but we drop them for clarity.

\subsubsection{Telescope Error}
\label{sec:telescope}

Besides radiometer noise discussed previously, additional noise contributions from the telescope or its backends or receivers, $\sigma_{\tel}$, such as from RFI, will contribute to the overall TOA uncertainty. Narrow-band RFI such as from artificial satellites will contaminate particular frequency channels in pulsar data that can be excised, making the effective bandwidth observed smaller without introducing additional TOA uncertainties, which we do not account for in our analysis. However, any remaining RFI will increase the TOA uncertainty.

Pulse shape changes from incorrect absolute gain calibration and summation of the polarization profiles into intensity profiles lead to additional TOA uncertainties from typical template fitting \citep{NG9WN}. For circularly polarized channels, the TOA uncertainty from gain variation is \added{(\citealt{Cordes+2004}; also see \citealt{Stinebring+1984} for an in-depth overview of alterations to the polarization properties of pulse profiles\addedA{; here we consider Gaussian profiles propagated through the Mueller matrices in Appendix A of that paper})}
\be
\sigma_{\rm pol} \sim 1~\mathrm{\mu s}~\left(\frac{\varepsilon}{0.1}\right)\left(\frac{\pi_{\rm V}}{0.1}\right)\left(\frac{W}{100~\mathrm{\mu s}}\right),
\label{eq:sigma_pol}
\ee
where $\varepsilon = \delta g/g$ is the fractional gain error, $\pi_{\rm V}$ is the degree of circular polarization, and $W$ is the pulse width. The fiducial values listed for the polarization parameters match closely to the NANOGrav AO data in the 1.4~GHz frequency band ($\varepsilon \sim 0.08$, $\pi_{\rm V}\sim 0.1$; P.\,A.\,Gentile et al. \added{submitted}). The uncertainty $\sigma_{\rm pol}$ can be a systematic or random error depending on how $\varepsilon$ varies in time. All three quantities are frequency dependent and therefore $\sigma_{\rm pol}$ is also a function of frequency. Instrumental self-polarization can also induce TOA uncertainties. Cross coupling for a circularly-polarized feed will result in a measured false circular polarization $\pi_{\rm V} \simeq 2 \eta^{1/2} \pi_{\rm L}$ assuming the receiver polarizations are orthogonal, where $\eta$ is the voltage cross-coupling coefficient and $\pi_{\rm L}$ is the degree of linear polarization for the pulsar \addedC{\citep[see again][]{Cordes+2004}}. The false circular polarization yields a TOA error as in Eq.~\ref{eq:sigma_pol}. Cross-coupling for our current set of receivers has not been well-measured and so we exclude the contribution to the TOA uncertainty in this work.

As with the various white-noise components to the timing uncertainty, the total uncertainty from polarization errors can be determined from the covariance matrix
\be
\Pmat = \mathrm{diag}\left[\sigma_{\rm pol}^2(\nu)\right]
\ee
and using Eq.~\ref{eq:epoch_average} to find the single value (squared) for the observation. Since telescope environments and instrumentation can vary wildly, we do not account for other telescope errors in our specific pulsar analysis though in general it will add into the total TOA uncertainty (Eq.~\ref{eq:sigma_TOA}).


\subsection{Integration Time Considerations}
\label{sec:time_effects}

Increased integration time decreases the white noise components of the TOA uncertainty; most NANOGrav pulsars are dominated by white noise and see large improvements in GW sensitivity by increased time per source \citep{NG9EN}. Increased observing time improves the average S/N by $T^{1/2}$ as per Eq~\ref{eq:SN}. The jitter error also improves by the same factor because the number of pulses averaged increases \citep{NG9WN}. The scintillation error scales according to Eq~\ref{eq:niss} and \ref{eq:DISS_error}, also as $\sim\!\!T^{1/2}$ when the number of scintles is large.

Increased integration time will also affect the number of scintles measured in a single observation. While the average observed flux across epochs remains the same (see Eq.~\ref{eq:sn_pdf}), the distribution will change; in the low-scintle limit the median S/N is much lower than the mean and the majority of measured TOA uncertainties will be below the mean TOA uncertainty. Since many MSPs have white-noise TOA uncertainties dominated by template-fitting errors, ensuring that multiple scintles are observed every epoch is imperative for robust timing-model fitting. \added{If the scintle size is of order the integration/bandwidth of the observation, the intensity will often be suppressed as per the PDF in Eq.~\ref{eq:sn_pdf}.} In the very low-S/N limit, determination of a TOA is not possible at all (again see Appendix B of \citealt{NG9yr}).

In the case where multiple frequency bands are observed non-simultaneously within an epoch, an additional uncertainty due to time-variable DM must be accounted for in the DM estimation process \citep{Lam+2015}. While typically a minor effect on the TOA uncertainty, the error grows when the time between observation increases, when $\Dtd$ is small, or when the DM over the line-of-sight rapidly changes such as when a pulsar is observed close to the Sun \added{\citep{y+07b,hsd+2016,NG9DM}}. For optimal DM correction and reduced TOA uncertainty, observing with a wide band simultaneously is preferred.

\begin{figure}[t!]
\includegraphics[width=0.5\textwidth]{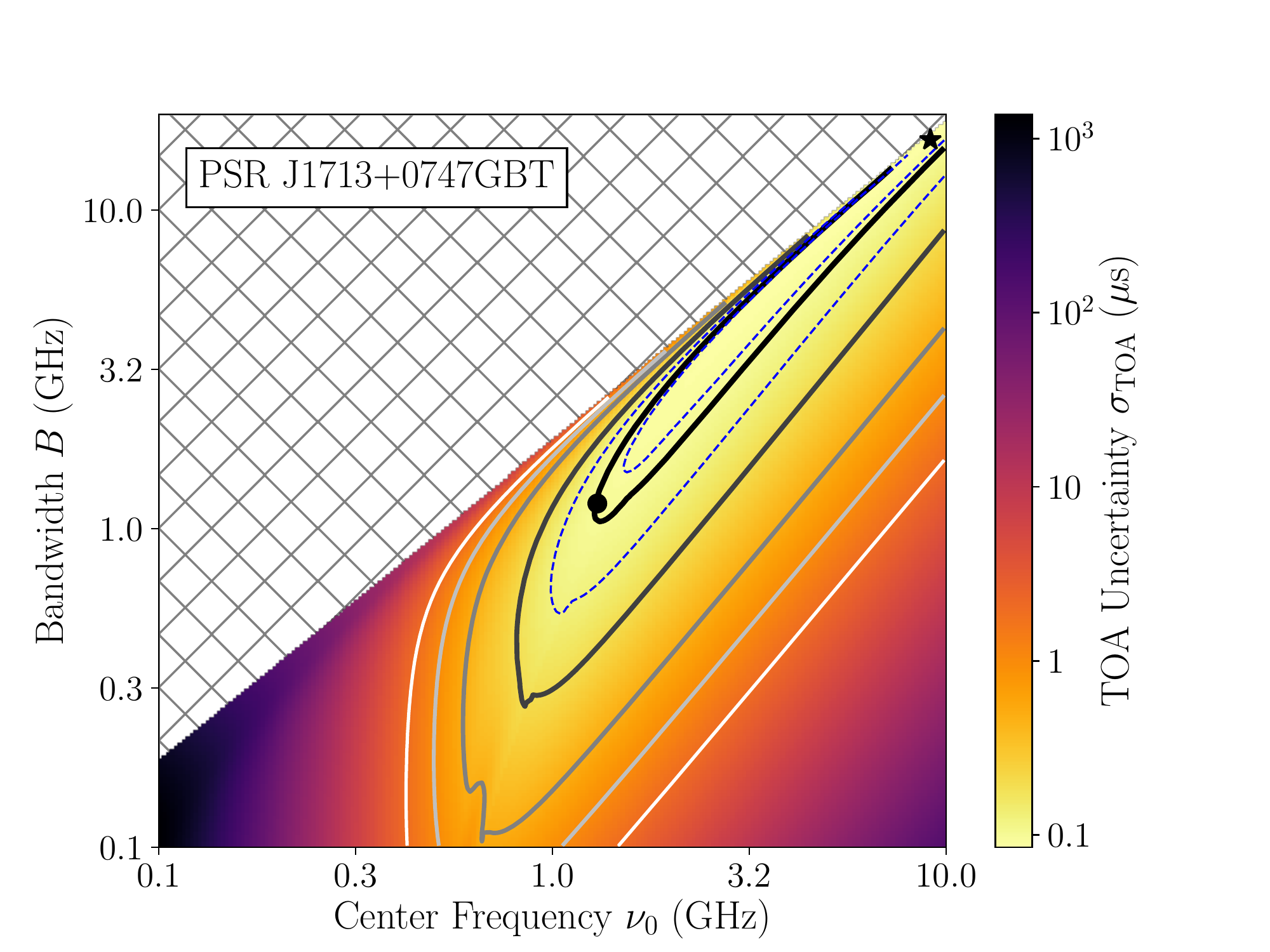}
  \caption{\footnotesize TOA uncertainty as a function of observing center frequency and bandwidth for PSR~J1713+0747 observed with the GBT. Pulsar and observation parameter values are provided in Table~\ref{table:parameters}. \addedC{Solid} contours indicate TOA uncertainties of \addedC{2, 1, 0.5, 0.2, and 0.1}~$\mathrm{\mu s}$, in order of increasing darkness and thickness. \addedC{The two dashed blue contours represent a 10\% and 50\% increase above the minimum $\sigma_{\rm TOA}$.} The hatched region represents where $B > 1.9\nu_0$. The distortion in the contours is an artifact in how we approximate $\Dmat$ (see Eq.~\ref{eq:Dmat}). The star identifies where the lowest possible TOA uncertainty occurs, \addedC{$\sigma_{\rm TOA}(\nu_0=9.1~\mathrm{GHz},B=16.6~\mathrm{GHz}) = 85~\mathrm{ns}$}. The circle indicates the approximate $\nu_0$ and $B$ used in current observations at the GBT, with $\sigma_{\rm TOA}(\nu_0=1.3~\mathrm{GHz},B=1.2~\mathrm{GHz}) = 99~\mathrm{ns}$. However, because NANOGrav currently uses two observing bands separately to cover the whole frequency range, it takes 60 minutes of on-sky time to obtain 30 minutes of integration across the band; therefore, the overall $\sigma_{\rm TOA}$ can be improved by a factor of $\sim\!\!\sqrt{2}$ using a single wideband receiver with the same integration time.}
\vspace{3ex}
\label{fig:J1713+0747GBT}
\end{figure}

\begin{center}
\begin{deluxetable*}{ccccccl}
\tablecolumns{7}
\tablecaption{Values Adopted for Specific NANOGrav MSPs$^{\rm a}$}
\tablehead{
\colhead{Pulsar} & \multicolumn{5}{c}{Pulsars} & \colhead{Notes}\\
\colhead{Parameters} & \colhead{J1713+0747} & \colhead{J1909$-$3744} & \colhead{J1903+0327} & \colhead{J1744$-$1134} & \colhead{J1643$-$1224} & \colhead{}\\
\colhead{} & \colhead{\S\ref{sec:J1713}} & \colhead{\S\ref{sec:J1909}} & \colhead{\S\ref{sec:J1903}} & \colhead{\S\ref{sec:J1744}} & \colhead{\S\ref{sec:J1643}}
}
\startdata
Dispersion Measure & 15.97 & 10.39 & 297.52 & 3.14 & 62.41 & \citet{NG11yr}.\\
DM~(pc~cm$^{-3}$) & & & & & &\\
\hline
Flux Density $I_0$~(mJy) & 10.3 &  2.6 &  2.6 &  4.9 &  9.1 & Least-squares fit to flux density values \\
Flux spectral index $\alpha$ & $-1.20$ & $-1.89$ & $-2.08$ & $-1.49$ & $-2.23$ & provided in \citet{NG5yr} \\
& & & & & & and \citet{PSRCAT}.\\
\hline
Scintillation timescale & 1760$^{\rm b}$ &  1390$^{\rm b}$ & 7.4$^{\rm c}$ &  1270$^{\rm b}$ &  360$^{\rm b}$ & See footnotes.\\
 $\Dtdo$~(s) & & & & & &\\
Scattering timescale & 0.052$^{\rm d}$ &  0.028$^{\rm d}$ &  554$^{\rm e}$ & 0.026$^{\rm d}$ & 43$^{\rm b}$ & \\
 $\taudo$~($\mu\mathrm{s}$)  & & & & & &\\ 
Scattering timescale & 0.035 & 0.021 & 277 & 0.012  &  21.5 & \\
variation $\delta \taudo$~($\mu\mathrm{s}$)$^{\rm f}$ & & & & & &\\
\hline
Signal-to-noise scaling & 26.4/26.8$^{\rm g}$ & 61.4 & 9.3 & 27.0 & 10.2 & Taken from measurements\\
factor $\bar{U}_{\rm obs}$ & & & & & & in the 1.4~GHz band from\\
Effective pulse width & 533/539$^{\rm g}$ &  261 &  405 &  511 & 973 & \citet{NG9WN}. \\
$\Weff~(\mathrm{\mu s}$) & & & & & & \\
Pulse width at half & 109/110$^{\rm g}$ &  41 & 195  &  137 & 315 & \\
maximum $W_{50}~(\mathrm{\mu s}$) & & & & & &\\
Rms jitter $\sigma_\J$~($\mu\mathrm{s}$) & 0.051/0.039$^{\rm g}$ & 0.014 &  0.257 &  0.066 &  0.219 & \\
& & & & & &\\
\hline
\hline
Telescope & & & & & &\\
Parameters & & & & & &\\
\hline
Gain~(K/Jy) & 2/10$^{\rm g}$ & 2 & 10 & 2 & 2 & $A_e = 27600$ for AO, \\
 & & & & & & $5520$ for GBT
\enddata
\footnotetext{Values provided are referenced to 1~GHz. The Galactic background temperature is set to scale proportional to  $\nu^{-2.75}$. We set $T_{\rm const} = 30$~K, $\varepsilon = 0.08$, and $\pi_{\rm V} = 0.1$. We used an integration time of $T = 30$~minutes.}
\footnotetext{Scaled from \citet{Keith+2013}.}
\footnotetext{Estimated from \citet{NE2001}.}
\footnotetext{Scaled from \citet{Levin+2016} assuming $C_1 = 1.16$.}
\footnotetext{Scaled from \citet{Champion+2008}.}
\footnotetext{For an approximation of $t_C$, we used scattering timescale variations taken as half the maximum range in \citet{Levin+2016}, otherwise we assumed $\delta \taudo = 0.5\taudo$. See text for more details.}
\footnotetext{For PSR~J1713+0747, we list GBT and then AO parameters when two numbers are provided.}
\label{table:parameters}
\end{deluxetable*}
\end{center}

\section{Implications for NANOGrav}
\label{sec:NANOGravMSPs}

In this section, we describe our analysis of five specific MSPs from the NANOGrav PTA. \added{We considered pulsars observed with the two telescopes primarily used by NANOGrav: the Green Bank Telescope (GBT) at the Green Bank Observatory and the Arecibo Observatory (AO). Each telescope has multiple receivers to cover portions of the total frequency band, with observations at each band being nearly contiguous at AO but with multiple days of spacing at the GBT for observational efficiency. Most pulsars are observed over 2$-$3 frequency bands chosen to optimize the long-term timing rms and with a $\sim$30-day cadence though several are observed weekly \citep{NG11yr}.}

Our goal was to calculate $\sigma_{\rm TOA}(\nu_0,B)$ for a given set of pulsar- and telescope-specific parameters. Since we covered a large range in center frequencies and bandwidths, we chose to calculate over the logarithm of each quantity, i.e., $\sigma_{\rm TOA}(\log \nu_0,\log B)$, between 0.1 and 10.0~GHz. We excluded bandwidths greater than 1.9 times the center frequency from the analysis and set the maximum $\sigma_{\rm TOA}$ to be the pulsar spin periods. We assumed a constant gain of 2~K/Jy at GBT and 10~K/Jy at AO.

\begin{figure}[t!]
\includegraphics[width=0.5\textwidth]{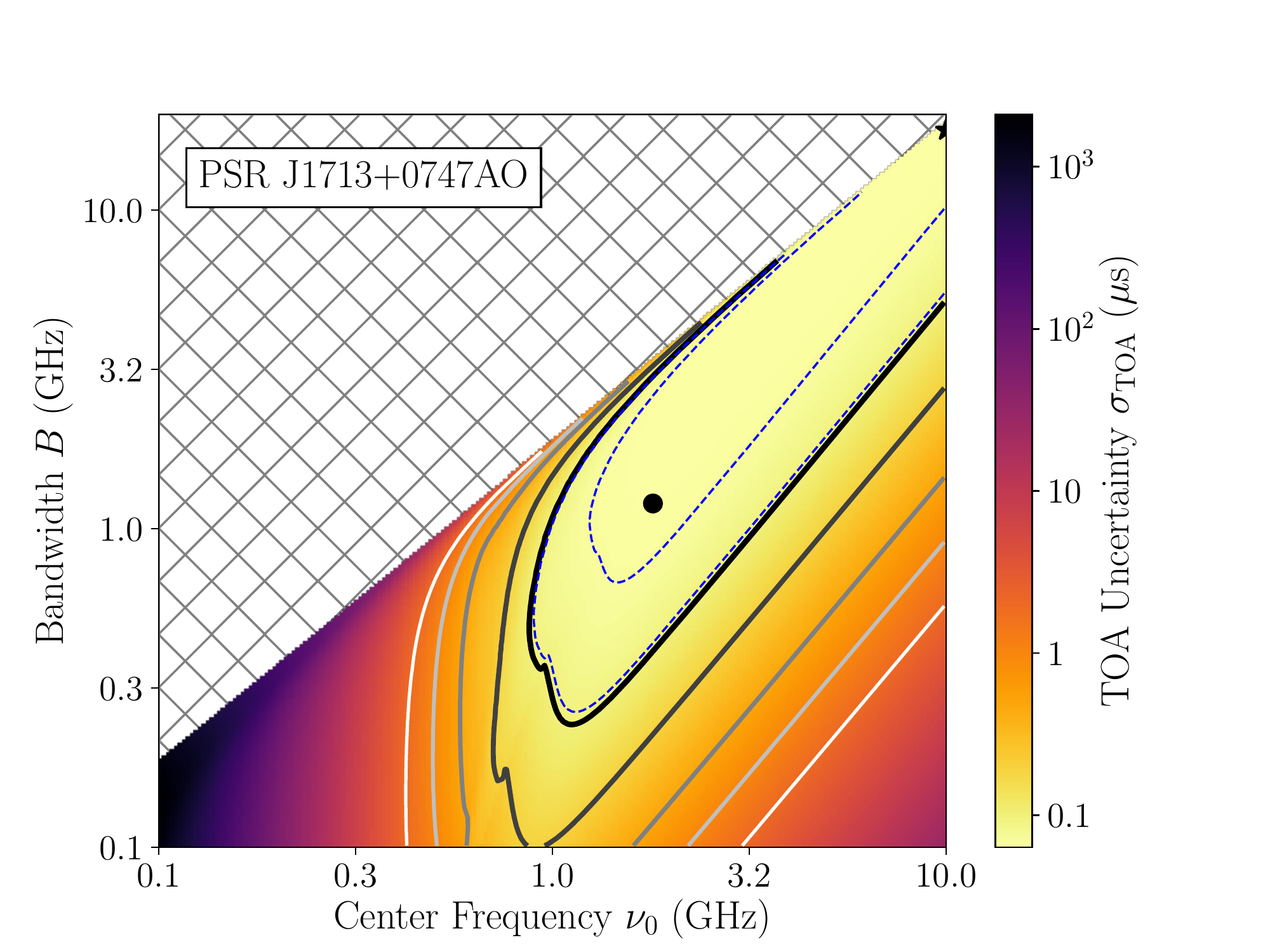}
  \caption{\footnotesize TOA uncertainty as a function of observing center frequency and bandwidth for PSR~J1713+0747 observed with AO. See Figure~\ref{fig:J1713+0747GBT} caption for more details. \addedC{Solid} contours \addedC{here again} indicate TOA uncertainties of \addedC{2, 1, 0.5, 0.2, and 0.1}~$\mathrm{\mu s}$, in order of increasing darkness and thickness. The minimum TOA uncertainty (black star) is $\sigma_{\rm TOA}(\nu_0=10.0~\mathrm{GHz},B=17.8~\mathrm{GHz}) = 63~\mathrm{ns}$ and the estimate given the current frequency coverage (black circle) is $\sigma_{\rm TOA}(\nu_0=1.8~\mathrm{GHz},B=1.2~\mathrm{GHz}) = 66~\mathrm{ns}$. 
}
\vspace{3ex}
\label{fig:J1713+0747AO}
\end{figure}

To simplify our calculations, we chose sub-band bandwidths such that there would be multiple scintles in each sub-band; this is not a common practice for pulsar timing observations. Such a choice also ensures that the pulse S/N would be approximately $S_0$ and the scintillation bandwidth would be approximately constant over the sub-band. At the highest frequencies however, typically of order a few GHz, scattering transitions from the strong to the weak regime \citep{Rickett1990,handbook}. In the weak scattering regime where DISS is attenuated by averaging effects, the PDF of the scintillation gain $g$ will be log-normally distributed rather than be in the $\chi^2$-form given in Eq.~\ref{eq:sn_pdf} (J.~M.~Cordes et al. in prep). Therefore, even at high frequencies when we would previously consider $\niss \sim 1$, the pulse S/N will tend towards the mean $S_0$. \added{However, to simplify our analysis, we ignore larger epoch-to-epoch variations in the S/N, noting that TOA uncertainties will be better on certain epochs and worse for others; a fuller analysis will want to account for such a variation.} We note that for some lines of sight such as those through the Galactic plane, scattering will remain in the strong regime beyond our frequency limit of 10~GHz, with the largest effects in directions close to the Galactic center.

Since scintillation noise also depends on the number of scintles in the band but changes between the strong and weak scattering regime, we approximate $\Dmat$ as follows. When there are multiple scintles ($\niss \ge 2$) in the frequency sub-band, we assume that the sub-bands are uncorrelated with each other. In the other case, we assume that the sub-bands are completely correlated ($\rho_\DISS = 1$) and therefore, the scintillation covariance matrix will take the block-diagonal form
\be
\Dmat \approx
\begin{bmatrix}
\mathrm{diag}\left[\sigma_{\DISS}^2(\nu|\Dtdo,\Dnudo)\right] & 0 \\
0 &  \sigma_{\DISS}(\nu|\Dtd,\Dnudo)\times\\
& \sigma_{\DISS}(\nu^\prime|\Dtd,\Dnudo)
\end{bmatrix}.
\label{eq:Dmat}
\ee
For our analysis, we chose 100 sub-bands for each calculation of $\sigma_{\rm TOA}(\nu_0,B)$.

Without prior knowledge of the amplitude of the extra chromatic delay $t_C(\nu)$, we used half the total variation (maximum minus minimum) in $\taudo$ given in \citet{Levin+2016} as an estimate for the amplitude of the chromatic delay away from the nominal scattering timescale. We set the frequency dependence proportional to \added{a fiducial value of} $\nu^{-4.4}$, allowing us to include this additional systematic uncertainty in our analysis. In cases where the variation was not measured, we used $\delta \taudo = 0.5\taudo$, which is roughly the amplitude for pulsars with measured variations.

Table~\ref{table:parameters} lists the parameters we included and the values we adopted for our selection of MSPs. For simplicity, we assumed that the telescope parameters were constant over all frequencies. In addition, we chose $\Weff$, $W_{50}$, and $\sigma_\J$ to be constant over all frequencies, using the 1.4~GHz values in \citet{NG9WN}. In reality, the values do vary across frequency but we do not have information at the lowest and highest frequencies in our analysis. Interpolating values over the frequency range we do have information for does not change our results substantially. In the next subsections, we discuss the results for the pulsars individually. The code that we used in our analysis is freely available at \url{https://github.com/mtlam/FrequencyOptimizer}\footnote{The version used in this paper is available as\\ submission\_version.tar.gz}.

\subsection{PSR~J1713+0747}
\label{sec:J1713}

The precision timing for PSR~J1713+0747 has led to its usage in a number of test of gravity and it typically dominates the sensitivity of PTAs towards gravitational waves \citep[e.g.,][]{Zhu+2015,NG5BWM}. Current observations of the pulsar by NANOGrav are conducted at both the GBT and AO with a weekly cadence because of its high precision timing. We therefore perform our calculations for this pulsar at both telescopes but separately.

\begin{figure}
\includegraphics[width=0.5\textwidth]{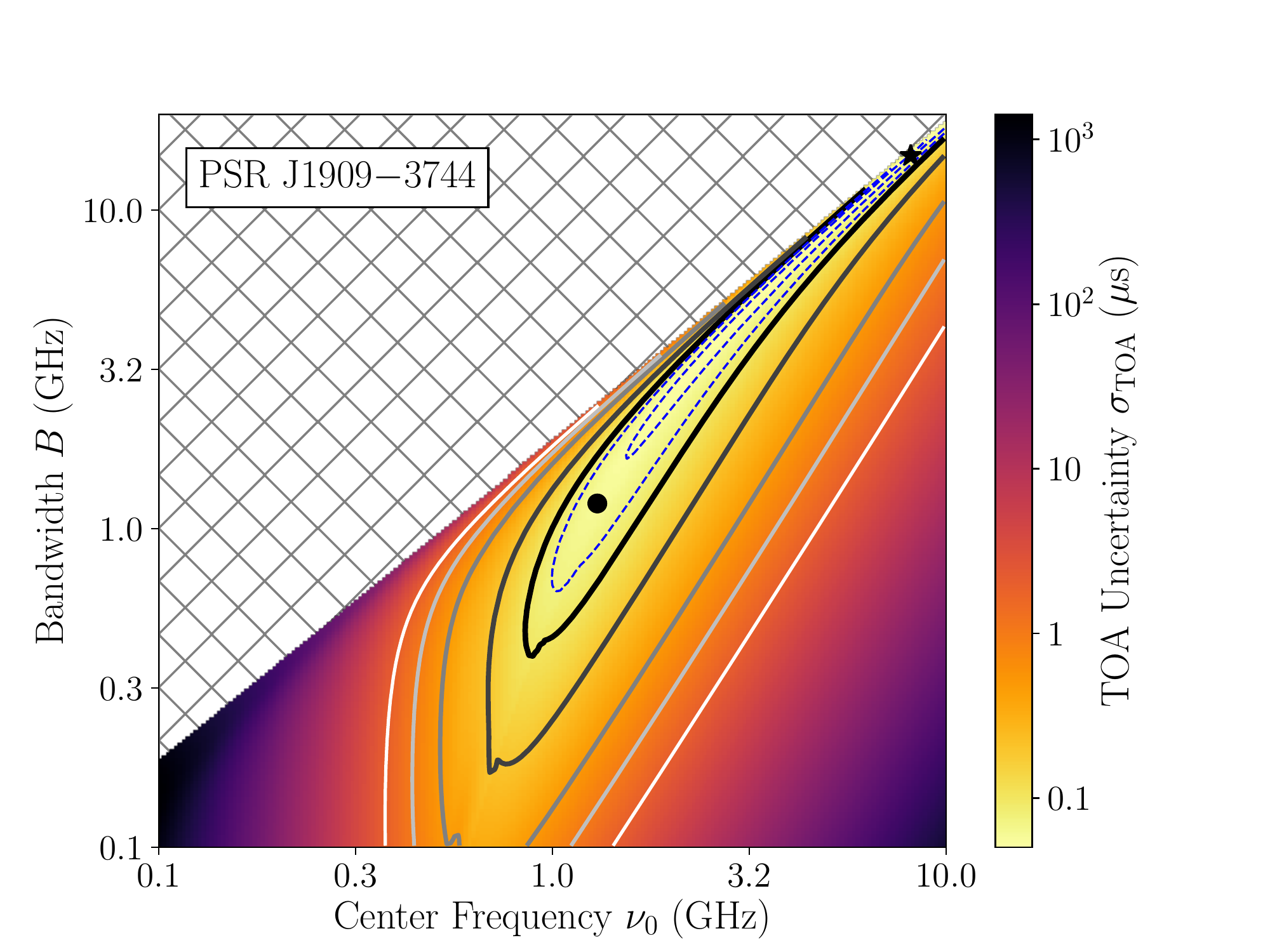}
  \caption{\footnotesize TOA uncertainty as a function of observing center frequency and bandwidth for PSR~J1909$-$3744 observed with the GBT. See Figure~\ref{fig:J1713+0747GBT} caption for more details. \addedC{Solid} contours indicate TOA uncertainties of \addedC{2, 1, 0.5, 0.2, and 0.1}~$\mathrm{\mu s}$, in order of increasing darkness and thickness. The minimum TOA uncertainty (black star) is $\sigma_{\rm TOA}(\nu_0=8.1~\mathrm{GHz},B=14.8~\mathrm{GHz}) = 50~\mathrm{ns}$ and the estimate given the current frequency coverage (black circle) is  $\sigma_{\rm TOA}(\nu_0=1.3~\mathrm{GHz},B=1.2~\mathrm{GHz}) = 59~\mathrm{ns}$.}
\label{fig:J1909-3744}
\end{figure}


Figure~\ref{fig:J1713+0747GBT} shows the results of our optimal frequency and bandwidth analysis for PSR~J1713+0747 for 30-minute observations at the GBT. The \addedC{solid} contours represent TOA uncertainties of \addedC{2, 1, 0.5, 0.2, and 0.1}~$\mathrm{\mu s}$, in order of increasing darkness and thickness. \addedC{The two dashed blue contours show the 10\% and 50\% increase above the minimum $\sigma_{\rm TOA}$.} The hatched region represents where $B > 1.9\nu_0$. We remind the reader that in our calculations for systematic delays, we only correct for a DM term and no scattering estimation is performed, thus mimicking current observing schemes. The increase in TOA uncertainty at the highest bandwidths comes partially from the lack of scattering correction (though a full scattering correction still yields a scattering misestimation term, for more see Figure 9 of \citealt{cs2010}) and partially from an increased contribution from frequency-dependent DM. If epoch-to-epoch DM variations were not removed, the TOA uncertainties will drastically increase towards larger bandwidths as well.


The black star in Figure~\ref{fig:J1713+0747GBT} shows the minimum TOA uncertainty is $\sigma_{\rm TOA}(\nu_0=9.1~\mathrm{GHz},B=16.6~\mathrm{GHz}) = 85~\mathrm{ns}$. The black circle shows NANOGrav's current frequency coverage assuming a continuous frequency range from $\sim$0.7 to $\sim$1.9~GHz, i.e., a center frequency of $\sim$1.3 GHz and a bandwidth of $\sim$1.2~\added{GHz}. In reality, there is a slight gap in frequency coverage from $\sim$0.9 to 1.1~GHz and high cadence ($\sim$ weekly instead of monthly) observations only cover the higher frequencies above the gap \citep{NG11yr}, thus complicating the analysis. Ignoring gaps, the TOA uncertainty is $\sigma_{\rm TOA}(\nu_0=1.3~\mathrm{GHz},B=1.2~\mathrm{GHz}) = 99~\mathrm{ns}$.

The above analysis assumes the standard NANOGrav receiver setup, which requires 60~minutes of on-sky time in order to obtain 30~minutes of effective integration time uniformly across the entire band. Were a broadband receiver available, capable of covering the entire 0.7~GHz to 1.9~GHz band, the overall $\sigma_{\rm TOA}$ could be improved by a factor of $\sim\!\!\sqrt{2}$ using a single wideband receiver with the same effective integration time. \addedC{We see that for this pulsar and observing setup, we lie within the phase-space valley very close to the minimum for the average S/N epoch though some slight improvements can be made by going to higher frequencies.}



Figure~\ref{fig:J1713+0747AO} shows the same analysis but completed for AO. We see that because of the sensitivity of AO, the same TOA uncertainty covers a larger frequency\addedC{-bandwidth region}. An increase in the gain improves the pulse S/N overall, which allows for reduced TOA uncertainties at higher frequencies where the pulsar power-law spectral flux is diminished but also at lower frequencies where scattering begins to reduce the flux as well; our analysis demonstrates why considering every contribution to the TOA uncertainty is important. For PSR~J1713+0747 observed with AO, the TOA uncertainty at the minimum is $\sigma_{\rm TOA}(\nu_0=10.0~\mathrm{GHz},B=17.8~\mathrm{GHz}) = 63~\mathrm{ns}$.




At AO, NANOGrav observes PSR~J1713+0747 between $\sim$1.2 and $\sim$2.4~GHz, i.e., $\nu_0 = 1.8$~GHz and $B = 1.2$~GHz. The estimate of the TOA uncertainty for these parameters is $\sigma_{\rm TOA}(\nu_0=1.8~\mathrm{GHz},B=1.2~\mathrm{GHz}) = 66~\mathrm{ns}$. We note that since NANOGrav observes PSR~J1713+0747 with both telescopes, determining the combined optimal frequency ranges becomes more complex since the telescope gains and $T_{\rm const}$ values will be different, though the TOA uncertainty will be weighted more towards AO's improved sensitivity.

\begin{figure}
\includegraphics[width=0.5\textwidth]{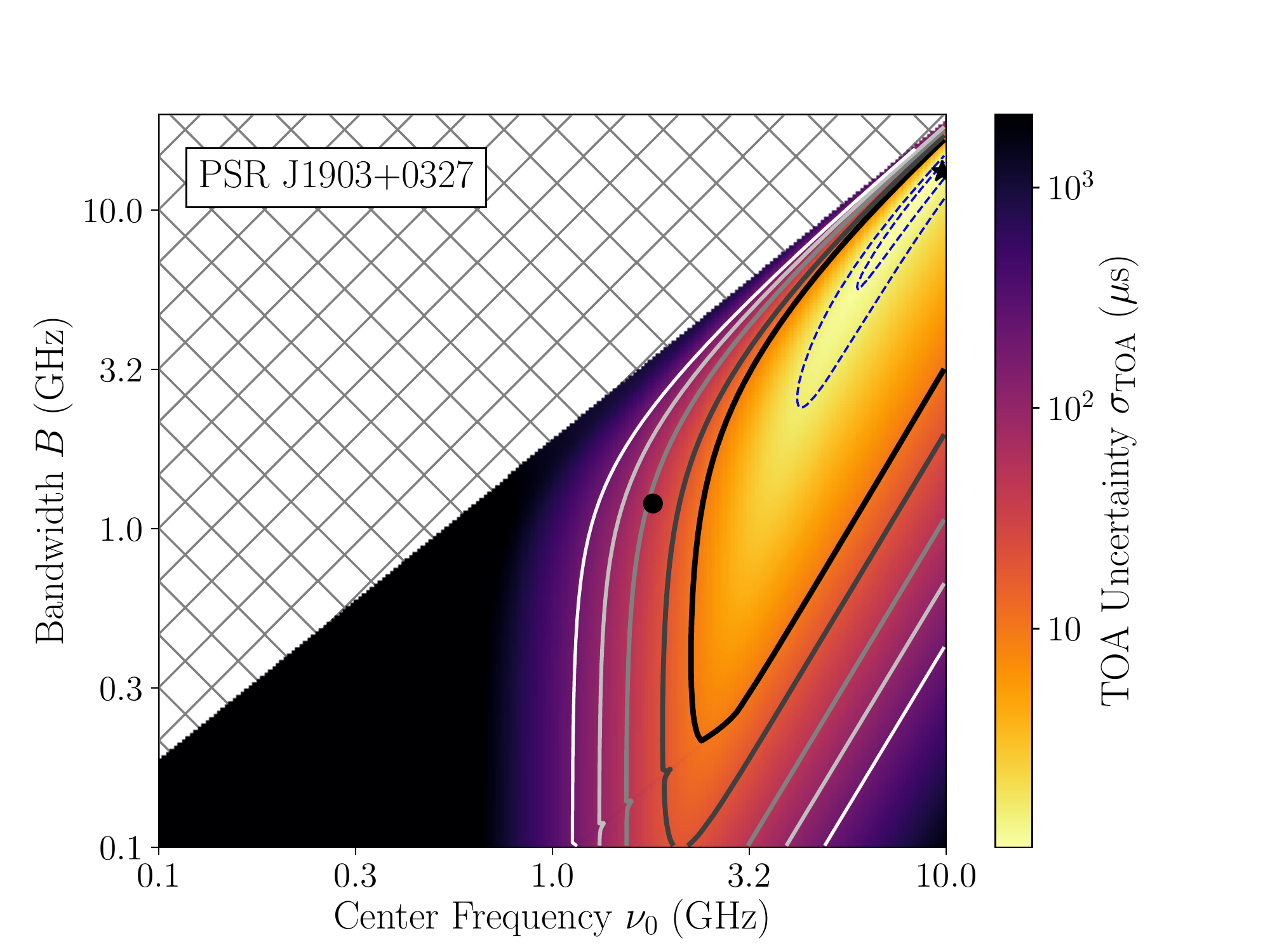}
  \caption{\footnotesize TOA uncertainty as a function of observing center frequency and bandwidth for PSR~J1903+0327 observed with AO. See Figure~\ref{fig:J1713+0747GBT} caption for more details. \addedC{Solid} contours indicate TOA uncertainties of \addedC{200, 100, 50, 20, and 10}~$\mathrm{\mu s}$, in order of increasing darkness and thickness. The minimum TOA uncertainty (black star) is $\nu_0=9.8~\mathrm{GHz},B=13.2~\mathrm{GHz} = 1.0~\mathrm{\mu s}$ and the estimate given the current frequency coverage (black circle) is  $\sigma_{\rm TOA}(\nu_0=1.8~\mathrm{GHz},B=1.2~\mathrm{GHz}) = 44.0~\mathrm{\mu s}$.}
\label{fig:J1903+0327}
\vspace{3ex}
\end{figure}

\subsection{PSR~J1909$-$3744}
\label{sec:J1909}

PSR~J1909$-$3744 is one of the best-timed pulsars. It has the second lowest amount of white noise and the lowest measured excess noise beyond the white noise \citep{NG9EN}. Therefore it is one of the most sensitive components in PTAs to low-frequency gravitational wave detection \citep[e.g.,][]{NG5CW,Shannon+2015}. Figure~\ref{fig:J1909-3744} shows the analysis for observations of PSR~J1909$-$3744 using the GBT. Again assuming the GBT parameters of a center frequency of $\sim$1.3~GHz and a bandwidth of $\sim$1.2~GHz, we see that we obtain similar results to PSR~J1713+0747 when observed with the GBT, although note \addedC{the narrowing of the optimal region in the phase space}. As reported for PSR~J1713+0747, the minimum TOA uncertainty for PSR~J1909$-$3744 is $\sigma_{\rm TOA}(\nu_0=8.1~\mathrm{GHz},B=14.8~\mathrm{GHz}) = 50~\mathrm{ns}$ and the estimate given the current frequency coverage is  $\sigma_{\rm TOA}(\nu_0=1.3~\mathrm{GHz},B=1.2~\mathrm{GHz}) = 59~\mathrm{ns}$.

\subsection{PSR~J1903+0327}
\label{sec:J1903}

\begin{figure}[t!]
\includegraphics[width=0.5\textwidth]{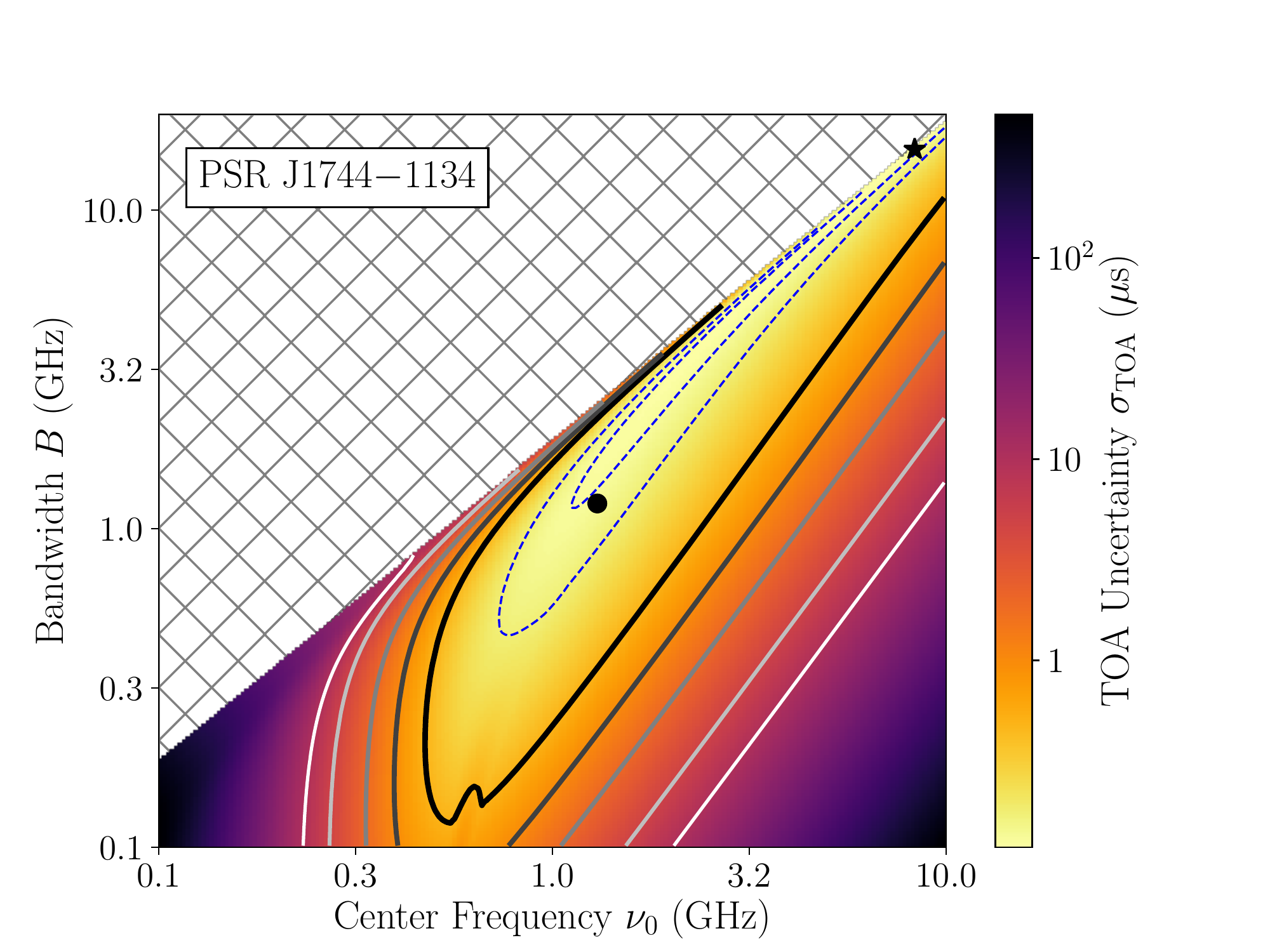}
  \caption{\footnotesize TOA uncertainty as a function of observing center frequency and bandwidth for PSR~J1744$-$1134 observed with the GBT. See Figure~\ref{fig:J1713+0747GBT} caption for more details. \addedC{Solid} contours indicate TOA uncertainties of \addedC{10, 5, 2, 1, and 0.5}~$\mathrm{\mu s}$, in order of increasing darkness and thickness. The minimum TOA uncertainty (black star) is $\sigma_{\rm TOA}(\nu_0=8.3~\mathrm{GHz},B=15.5~\mathrm{GHz}) = 120~\mathrm{ns}$ and the estimate given the current frequency coverage (black circle) is  $\sigma_{\rm TOA}(\nu_0=1.3~\mathrm{GHz},B=1.2~\mathrm{GHz}) = 140~\mathrm{ns}$.}
\label{fig:J1744-1134}
\end{figure}

PSR~J1903+0327 is the pulsar with the highest DM (297.52~pc~cm$^{-3}$) currently observed by NANOGrav. It shows considerable long-term red noise, both frequency-dependent and independent, suggesting unmodeled ISM and possibly intrinsic effects \citep{NG9EN}. Figure~\ref{fig:J1903+0327} shows the analysis for observations of PSR~J1903+0327 using AO. Like PSR~J1713+0747 observed with AO, NANOGrav observes this pulsar with $\nu_0 = 1.8$~GHz and $B = 1.2$~GHz, denoted by the black dot. 

The minimum TOA uncertainty is $\sigma_{\rm TOA}(\nu_0=9.8~\mathrm{GHz},B=13.2~\mathrm{GHz}) = 1.0~\mathrm{\mu s}$ and the estimate given the current frequency coverage is  $\sigma_{\rm TOA}(\nu_0=1.8~\mathrm{GHz},B=1.2~\mathrm{GHz}) = 44.0~\mathrm{\mu s}$. \addedC{Note that} since the Galactic latitude of the pulsar is $\approx -1^\circ$, the pulsar emission remains in the strong-scattering regime as discussed at the beginning of the section. Therefore, the S/N at high frequencies will be reduced for a majority of the epochs and the observed pulses will be scintillated high a small fraction of the time. For a typical epoch, the reduction in S/N implies that the center frequency will be pushed to lower frequencies where the S/N is higher. More work is needed to determine the effect of epoch-to-epoch variation in the scintillation gain on the long-term timing rms.


We see that the \addedC{current projected} TOA uncertainty is quite poor for this pulsar \addedC{and an order of magnitude larger than measured in previous work \citep{NG9EN,NG11yr}. The dominant factor for this discrepancy is the unknown scattering timescale variation parameter $\delta\taudo$. Since the variations (and the value itself) were unmeasured in \citet{Levin+2016} and we have adopted a fiducial value of 50\% of $\taudo$ when we cannot estimate them, we arrive at such a large number for $\sigma_{\rm TOA}$. If we reduce the variation to 10\%, the contours shift to the left some, such that we match with observed residual values. Many pulsars show variations of order a few in their scintillation parameters (see e.g., \citealt{Coles+2015} and \citealt{Levin+2016}) so we could not justify using such a low value. However, while we see that there is a large discrepancy, the overall trend in the phase space remains quite clear. For this pulsar, observing at higher frequencies and bandwidths will improve its timing quality.} Scattering corrections may be able to improve the overall timing of this pulsar \addedC{even further, as well as} others with high DM \citep{sc2017}.


\begin{figure}[t!]
\includegraphics[width=0.5\textwidth]{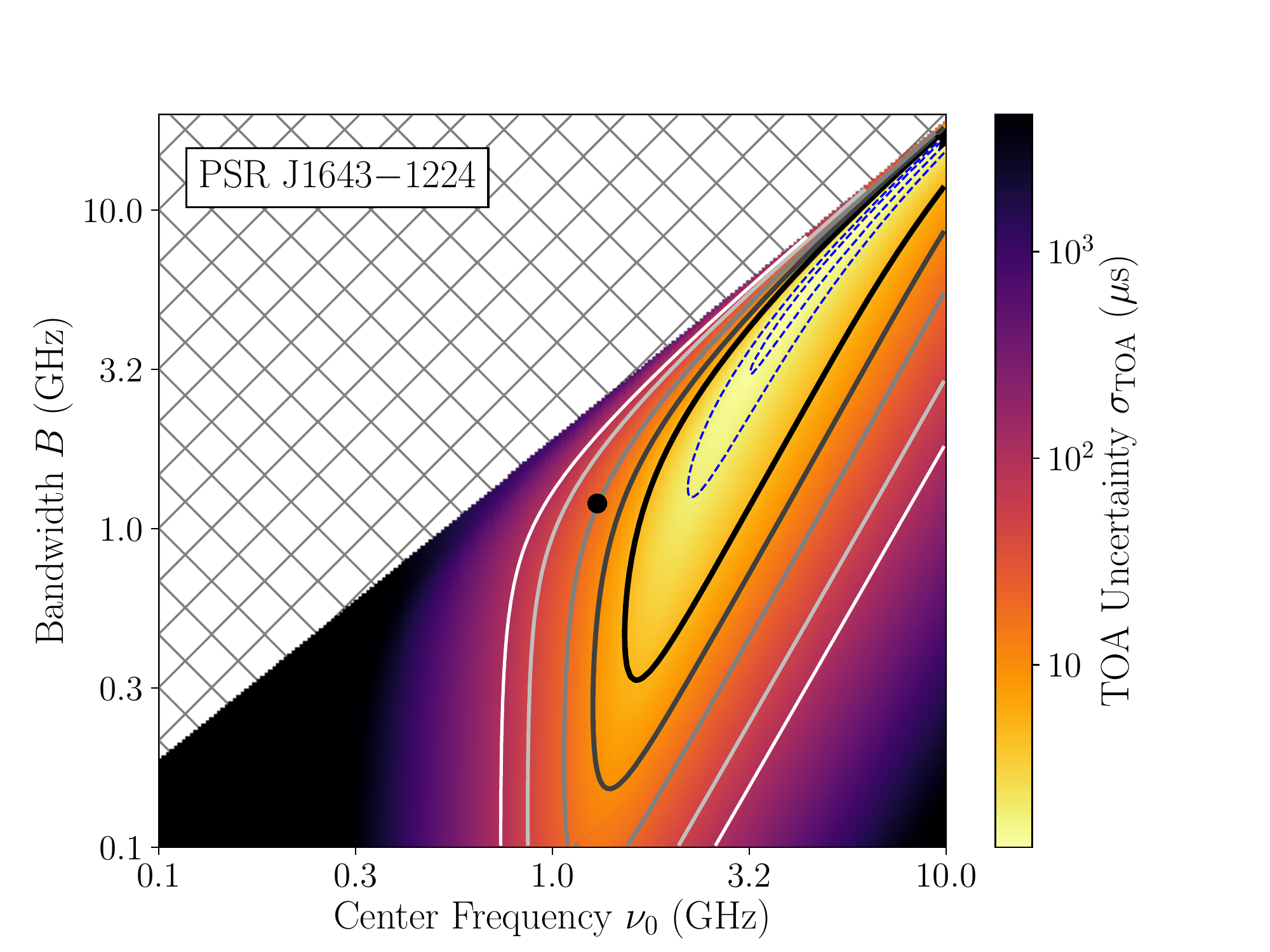}
  \caption{\footnotesize TOA uncertainty as a function of observing center frequency and bandwidth for PSR~J1643$-$1224 observed with the GBT. See Figure~\ref{fig:J1713+0747GBT} caption for more details. \addedC{Solid} contours indicate TOA uncertainties of \addedC{10, 5, 2, 1, and 0.5}~$\mathrm{\mu s}$, in order of increasing darkness and thickness. The minimum TOA uncertainty (black star) is $\sigma_{\rm TOA}(\nu_0=10.0~\mathrm{GHz},B=16.6~\mathrm{GHz}) = 1.3~\mathrm{\mu s}$ and the estimate given the current frequency coverage (black circle) is  $\sigma_{\rm TOA}(\nu_0=1.3~\mathrm{GHz},B=1.2~\mathrm{GHz}) = 20.8~\mathrm{\mu s}$.}
\label{fig:J1643-1224}
\end{figure}

\subsection{PSR~J1744$-$1134}
\label{sec:J1744}

On the other extreme, PSR~J1744$-$1134 is the pulsar with the lowest DM (3.09~pc~cm$^{-3}$) currently observed by NANOGrav. We show it as an example of a ``typical'' NANOGrav MSP with TOAs spanning the full length of our observing campaign and with scintillation parameters and rms timing residuals (TOAs minus timing model) comparable to many other NANOGrav pulsars. It does show frequency-independent excess white noise beyond the white noise in its timing residuals though no detectable red noise over the first eleven years of observation \citep{NG9EN,NG11yr}.

Figure~\ref{fig:J1744-1134} shows the analysis for observations of PSR~J1744$-$1134 using the GBT. The frequency coverage is the same as for PSR~J1909$-$3744. Note that at the highest bandwidths, we see the best TOA uncertainties for a given center frequency because the various delay misestimation terms do not dominate for a pulsar with such a low DM unlike in the previous examples. Therefore, observing with the highest possible bandwidths becomes important for pulsars similar to PSR~J1744$-$1134, many of which are currently being observed in PTA efforts. The minimum TOA uncertainty is $\sigma_{\rm TOA}(\nu_0=8.3~\mathrm{GHz},B=15.5~\mathrm{GHz}) = 120~\mathrm{ns}$ and the estimate given the current frequency coverage is  $\sigma_{\rm TOA}(\nu_0=1.3~\mathrm{GHz},B=1.2~\mathrm{GHz}) = 140~\mathrm{ns}$.

\subsection{PSR~J1643$-$1224}
\label{sec:J1643}

PSR~J1643$-$1224 is a high-DM pulsar (62.41~pc~cm$^{-3}$) with evidence of red noise in the timing residuals possibly from uncorrected ISM effects \citep{IPTADR1noise,NG9EN,NG11yr}; the pulsar lies beyond the HII region Sharpless 2-27 (Gum 73) and thus plasma along the line of sight may be affecting the timing noise \added{along with the determination of pulsar parameters} \citep{Blitz+1982,HIPPARCOS,NG9PX,Reardon+2016}. Corrections \addedB{of interstellar scattering delays} can improve the timing of the pulsar \citep{Lentati+2017}. The pulsar has also shown variations in the pulse profile interpreted as a change in the magnetosphere and such variations can also cause deviations from the timing model \citep{slk+2016}.

Figure~\ref{fig:J1643-1224} shows the analysis for observations of PSR~J1643$-$1224 using the GBT. The frequency coverage is the same as the other GBT pulsars. \addedC{As with PSR~J1903+0327,} at the current center frequency, NANOGrav's timing noise is sub-optimal and would benefit from an increase in the center frequency. We stress that observation of pulse profiles at large bandwidths allows for estimation of other parameters, such as the scintillation timescale and bandwidth, and therefore it is important to collect as much data as possible and then choose what data will go into a timing analysis. As stated previously though, increases in the center frequency and the bandwidth jointly will lead to an improvement in $\sigma_{\rm TOA}$; such increases become important for pulsars like PSR~J1643$-$1224 with moderate DMs and chromatic noise. For PSR~J1643$-$1224, the minimum TOA uncertainty is $\sigma_{\rm TOA}(\nu_0=10.0~\mathrm{GHz},B=16.6~\mathrm{GHz}) = 1.3~\mathrm{\mu s}$ \addedC{(recall that we strict our analysis to $\nu_0 \leq 10.0~\mathrm{GHz}$)} and the estimate given the current frequency coverage is  $\sigma_{\rm TOA}(\nu_0=1.3~\mathrm{GHz},B=1.2~\mathrm{GHz}) = 20.8~\mathrm{\mu s}$. \addedC{The unknown scattering timescale variation $\delta\taudo$ for this high DM pulsar is again the biggest contributor to the difference between the estimated current $\sigma_{\rm TOA}$ and the observed values in the literature.}

\begin{figure}[t!]
\includegraphics[width=0.5\textwidth]{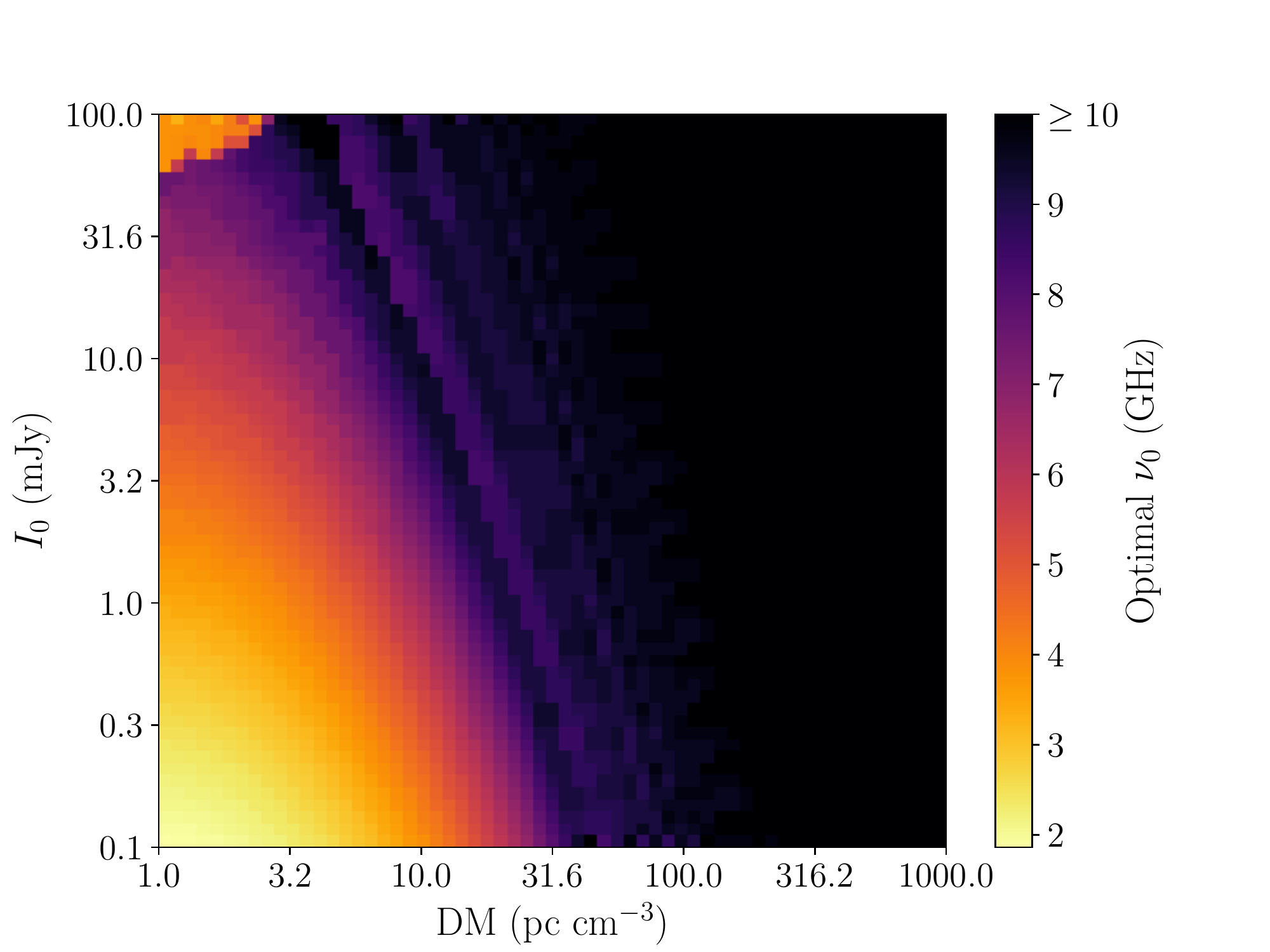}
  \caption{\footnotesize Optimal center frequency $\nu_0$ as a function of flux density at 1~GHz $I_0$ and DM for pulsar parameters discussed in \S\ref{sec:broadimplications}. Here we allow for any possible bandwidth. The black region denotes where \addedB{$\nu_0 \ge 10$~GHz} and is thus excluded from our analysis. The bands in the middle of the plot are a result of the minimum $\sigma_{\rm TOA}$ moving near our bandwidth cutoff $B = 1.9\nu_0$. The minimum optimal $\nu_0$ of this phase space is \addedC{around 1.9~GHz}. \addedC{The ``ripple'' feature around the highest frequencies results from the minimum $\sigma_{\rm TOA}$ wandering around the $B = 1.9\nu_0$ cutoff. The feature in the top left represents slight numerical wandering of the minimum $\sigma_{\rm TOA}$ in a large valley in the $\nu_0,B$ phase space and is over-represented in area in this figure due to the logarithmic axes.}}
\vspace{1ex}
\label{fig:phasespace}
\end{figure}

\section{Broad Implications for Precision Pulsar Timing}
\label{sec:broadimplications}

We have seen in the previous section our frequency optimization analysis for five example pulsars. Now we extend our analysis broadly to more MSPs. Since the pulsar flux $I_0$ and DM are the most dominant parameters, we chose to vary these parameters while holding all other parameters fixed for a fiducial pulsar.

We assumed a spectral index \addedA{$\alpha = -1.6$}, a constant effective pulse width $\Weff = 500~\mu\mathrm{s}$, \addedC{$\bar{U}_{\rm obs} = 20$,} and rms jitter $\sigma_{\J,30~\mathrm{min}} = 100~\mathrm{ns}$. We also assumed that the observations are performed with AO. We saw slightly lower optimal center frequencies when using the GBT gain of 2~K/Jy but do not show the plots since they are qualitatively similar. \addedB{Note this trend is different from our results for PSR~J1713+0747, though we discuss the possible reasons shortly}. For the scattering timescale, we used the scaling from \citet{Bhat+2004},
\ba
\log \tau_{\rm d,ms} &= &-6.46 + 0.154 \log\DM + 1.07 (\log\DM)^2\nonumber\\
&& - (3.86 \pm 0.16)\log\nu_{\rm GHz},
\ea
to vary $\taud$ and assumed $\delta\taudo = 0.5\taudo$, allowing us to input a chromatic term $t_C$ into the analysis. The maximum-to-minimum variation in $\taud$ found in \citet{Levin+2016} was typically found to be of this order. Assuming a uniform Kolmogorov medium, the scintillation timescale is \citep{cr98}
\ba
\hspace{-3ex}& &\Dtdo \nonumber \\
\hspace{-3ex}& &=\left(\!\frac{V_{p,\mathrm{eff} \perp}}{2.53 \times 10^4~\mathrm{km/s}}\!\right)^{-1}\!\left[\!\left(\frac{D}{\mathrm{kpc}}\right)\!\!\left(\frac{\Dnudo}{\mathrm{MHz}}\right)\!\right]^{1/2}\!\!\left(\frac{\nu_0}{\mathrm{GHz}}\right)^{-1}~\!\!\mathrm{s}\nonumber \\
\hspace{-3ex}& & \approx 724~\mathrm{s}~\!\!\!\left(\!\frac{\mu_{\rm T}}{\mathrm{10~mas~yr^{-1}}}\!\right)^{\!\!-1}\!\left(\frac{D}{\mathrm{kpc}}\right)^{\!\!-1/2}\!\left(\!\frac{\taudo}{\mathrm{0.1~\mu s}}\!\right)^{\!\!-1/2}\!\left(\!\frac{\nu_0}{\mathrm{GHz}}\!\right)^{\!\!-1},~\phantom{\mathrm{aaa}}
\ea
where $V_{p,\mathrm{eff} \perp}$ is the pulsar's effective velocity perpendicular to the line of sight, which includes components from the pulsar and Earth velocities relative to their respective Galactic rotating frame and the random ISM velocity. Assuming the pulsar velocity is the dominant component, we can convert the equation to use the total proper motion, $\mu_{\rm T}$. We assumed that $D = 1$~kpc and $\mu_{\rm T} = 10~\mathrm{mas~yr^{-1}}$ as fiducial values \citep{NG9PX}.

\begin{figure}[t!]
\includegraphics[width=0.5\textwidth]{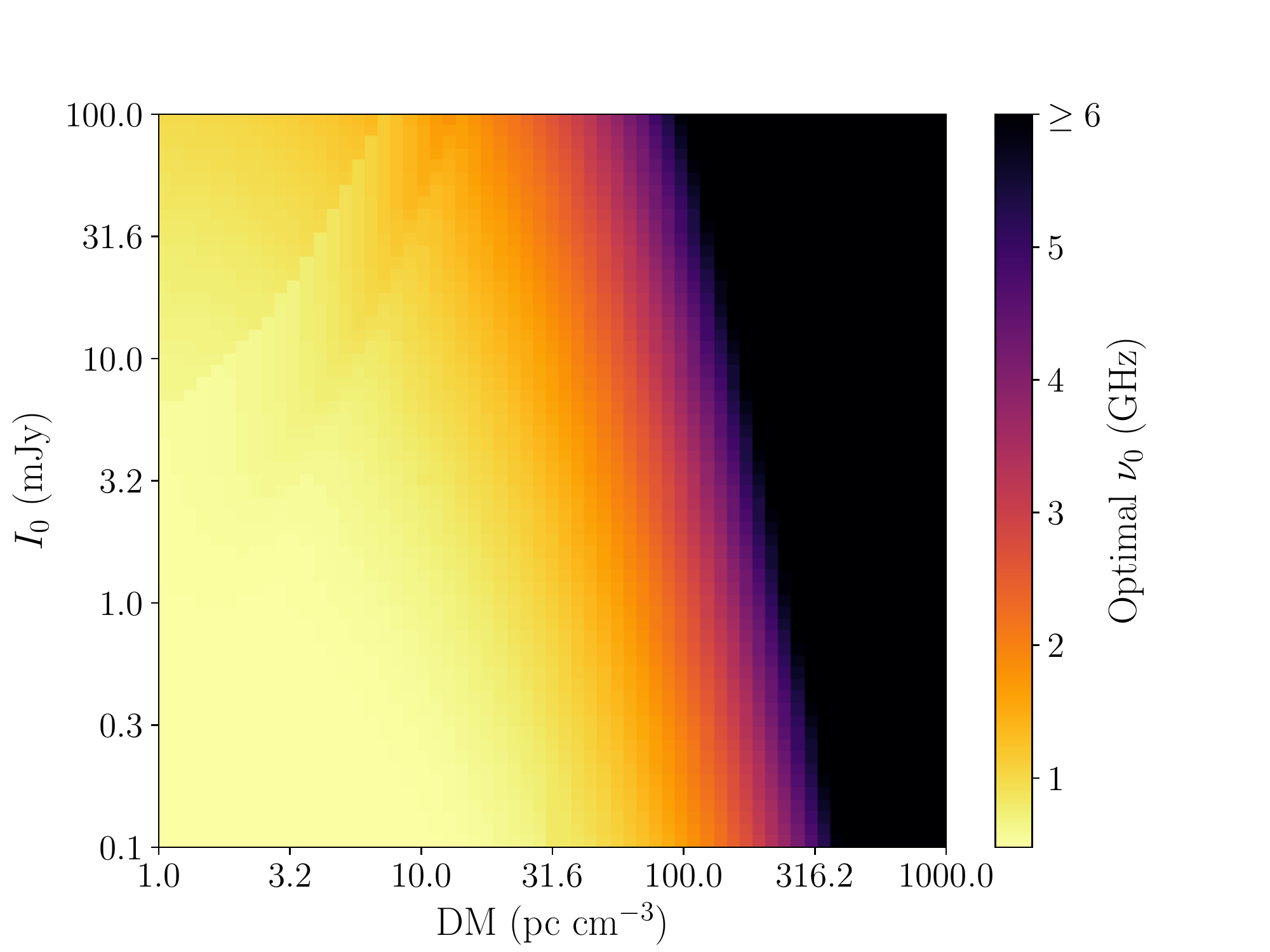}
  \caption{\footnotesize Optimal center frequency $\nu_0$ as a function of flux density at 1~GHz $I_0$ and DM for pulsar parameters discussed in \S\ref{sec:broadimplications}. Here we set the maximum allowed bandwidth to be half the center frequency ($B = 0.5\nu_0$) up to a bandwidth of 4~GHz. The black region denotes where \addedB{$\nu_0 \ge 6$~GHz} \addedC{for improved dynamic range in the plot}. The minimum optimal $\nu_0$ of this phase space is \addedA{around \addedC{0.48} GHz}. \addedC{As in Figure~\ref{fig:phasespace}, the ``ripple'' now in the top left represents the minimum $\sigma_{\rm TOA}$ wandering near the bandwidth cutoffs.} }
\label{fig:phasespace2}
\end{figure}


Figure~\ref{fig:phasespace} shows the optimal center frequency for our fiducial pulsar as a function of $I_0$ and DM allowing for any possible bandwidth. Pulsars with low DM, such as PSR~J1744$-$1134, obtain better $\sigma_{\rm TOA}$ when observed at lower center frequencies. Conversely, pulsars with high DM, such as PSR~J1903+0327, are dominated by ISM-related timing effects, which are partially mitigated by the higher center frequencies. \addedC{However, all $\nu_0$ were $\gtrsim1.9$~GHz, indicating the preference to observe with higher center frequencies than typical current observing configurations.}

Especially at larger optimal frequencies, we found typically that the optimal bandwidths were near or on the maximum of our allowed range ($B\le 1.9\nu_0$). Figure~\ref{fig:phasespace2} shows the optimal center frequency where we considered more realistic engineering constraints for a single receiver, with $B = 0.5\nu_0$ up to a maximum bandwidth of 4~GHz. We see that because of the constraints, lower optimal frequencies are preferred and are limited by the bandwidth restrictions. We stress that the bandwidth restrictions are not hard limitations but serve as approximate guidelines that may be improved upon in the future. As is current procedure, multiple telescopes and/or receivers can be used to observe a source within a single epoch and therefore the optimal bandwidths can be larger than what a single receiver will allow.

Increases in telescope collecting area will improve the pulse S/N and therefore fainter pulsars can be observed with similar $\sigma_{\rm TOA}$ as brighter pulsars. Since the S/N of pulses will improve with collecting area (i.e., the telescope gain), larger telescopes will have optimal center frequencies that may be preferentially higher than their smaller counterparts for objects with higher DMs but may in fact be lower for local objects with very low scattering. For any pulsar, observations that combine data from a variety of different telescopes can be tuned to cover the optimal frequency range for a given timing analysis and therefore provide the highest TOA precision possible.


\section{Discussion}
\label{sec:discussion}

For the first time, we have demonstrated a comprehensive methodology for calculating TOA uncertainties as a function of center frequency and bandwidth and therefore have determined the optimal frequencies for a pulsar to be observed given a set of pulsar-, Galactic-, and telescope-dependent parameters. So far we have only discussed the uncertainty on a TOA for a single epoch; an analysis of the long-term rms of the timing residuals must include additional noise terms that have not yet been fully accounted for, including systematic timing changes from temporal variations in parameters such as $t_C$. Our optimization analysis is very generic in its applications and can be used in any study that requires high-precision pulsar timing, including: tests of gravity, experiments in nuclear physics, determination of astrometry and stellar dynamics, detection of gravitational waves, probes of the ISM, and others.




Looking at the NANOGrav pulsars examined here, we \addedB{conclude that \addedC{many} MSPs timed in PTA observing programs \addedC{(e.g, the European Pulsar Timing Array and Parkes Pulsar Timing Array collaborations, see \citealt{Desvignes+2016} and \citealt{Reardon+2016}, respectively)} are done so with non-optimal frequency-bandwidth combinations. The range of optimality is much larger when observations are radiometer-noise-limited however. \addedC{For many MSPs, especially those at higher DMs}, PTA observations} would highly benefit from wideband systems typically centered at \addedC{somewhat} higher observing frequencies \addedC{than the receiver combinations used so far}. Very large increases in bandwidth can also help with ancillary data products that can provide additional information into the overall timing and noise model (e.g., in estimating per-epoch scintillation parameters); we do not suggest that wide-bandwidth systems be limited in size but rather that frequency ranges for should be carefully chosen for optimal TOA estimation specifically. Wide bandwidth systems can also help improve how optimal center frequencies are chosen and reduce the necessary integration time per pulsar used by current multiple-band systems, providing an instant reduction in the TOA uncertainty by $\sim\!\!\sqrt{2}$, assuming a fixed integration time.

While we have discussed how to optimize the frequency ranges for a given pulsar, optimization of an entire PTA requires consideration of additional factors. In general, the specific scientific goals will dictate the optimization of the array. For example, continuous gravitational wave sources benefit from observations of a few well-timed pulsars \citep{NG5CW}. A stochastic background of gravitational waves from sources such as supermassive black hole binaries or the inflationary epoch require many pulsars distributed across the sky \citep{NG9GWB}. In all cases, however, even small increases in TOA precision over many pulsars and many epochs can make a large difference in the detection and eventual characterization of GW sources.

In practice, implemented receiving systems tend to show a decrease in sensitivity (increase in system temperature) as the bandwidth increases above an octave.  This penalty has been only poorly quantified, however, and we have not tried to represent it in this work. However, this penalty would likely shift the optimal frequency slightly down and to the left in the $\sigma_{\rm TOA}(\nu_0,B)$ phase space if a larger than octave bandwidth is implemented. As new receivers and new telescopes become available, our formalism should be considered for the development of radio instrumentation used for high-precision pulsar timing.

\acknowledgments

{\it Acknowledgments.} \added{We thank the referees for a thorough reading of our manuscript and the improvements that resulted.} \addedC{The authors are members of the NANOGrav Physics Frontiers Center which is supported by NSF award number 1430284}. This work was supported in part by the ngVLA Community Studies program, coordinated by the National Radio Astronomy Observatory, which is a facility of the National Science Foundation operated under cooperative agreement by Associated Universities, Inc. In addition, part of this research was carried out at the Jet Propulsion Laboratory, California Institute of Technology, under a contract with the National Aeronautics and Space Administration.

\end{document}